\documentclass{article}
\usepackage{spconf,amsmath,epsfig,graphicx,color}
\graphicspath{{Eps/}}

\input{alphabet.tex}
\def\bdoc{\begin{document}}             \def\edoc{\end{document}}
\def\babs{\begin{abstract}}             \def\eabs{\end{abstract}}
\def\bcc{\begin{center}}                \def\ecc{\end{center}}
\def\ben{\begin{enumerate}}             \def\een{\end{enumerate}}
\def\bit{\begin{itemize}}               \def\eit{\end{itemize}}
\def\bpic{\begin{picture}}              \def\epic{\end{picture}}
\def\beq{\begin{equation}}              \def\eeq{\end{equation}}
\def\barr{\begin{array}}                \def\earr{\end{array}}
\def\btab{\begin{tabular}}              \def\etab{\end{tabular}}
\def\btabu{\begin{tabular}}             \def\etabu{\end{tabular}}
\def\bcc{\begin{center}}                \def\ecc{\end{center}}
\def\beqn{\begin{eqnarray}}             \def\eeqn{\end{eqnarray}}
\def\beqnx{\begin{eqnarray*}}           \def\eeqnx{\end{eqnarray*}}
\def\bfig{\begin{figure}}               \def\efig{\end{figure}}

\def\REM#1{}
\def\d#1{\,\mbox{d} #1}

\def\zerob{{\mathbf 0}}
\def\oneb{{\mathbf 1}} 

\def\intg{\int\kern-1.1em\int}
\def\expf#1{\exp\left[ {#1} \right]}
\def\cov#1{\mbox{cov}\left[ {#1} \right]}
\def\argmax#1#2{\arg\max_{#1}\left\{ #2 \right\}}
\def\argmin#1#2{\arg\min_{#1}\left\{ #2 \right\}}
\def\lra{\longrightarrow}

\def\gbu{\underline{\gb}}
\def\fbu{\underline{\fb}}
\def\zbu{\underline{\zb}}
\def\epsilonbu{\underline{\epsilonb}}
\def\thetabu{\underline{\thetab}}

\def\sigmae{{\sigma_{\epsilon}}}
\def\Sigmae{{\Sigmab_{\epsilonb}}}
\def\Sigmaf{{\Sigmab_{\fb}}}
\def\Sigmag{{\Sigmab_{\gb}}}

\def\fbuh{\widehat{\fbu}}
\def\zbuh{\widehat{\fbu}}
\def\thetabuh{\widehat{\thetabu}}

\def\fbh{\widehat{\fb}}
\def\Pbh{\widehat{\Pb}}
\def\Sigmabh{\widehat{\Sigmab}}
\def\Abh{\widehat{\Ab}}
\def\Bbh{\widehat{\Bb}}
\def\thetabh{\widehat{\thetab}}
\def\Rbe{\Rb_{\epsilon}}
\def\Rbeh{\widehat{\Rbe}}

\def\Rjk{{R_j}_k}
\def\fbik{{\fb_i}_k}
\def\fbjk{{\fb_j}_k}

\def\mik{{m_i}_k}
\def\sigmaik{{\sigma_i^2}_k}
\def\Sigmaik{{\Sigmab_i}_k}
\def\alphaik{{\alpha_i}_k}
\def\betaik{{\beta_i}_k}

\def\muik{{\mu_i}_k}
\def\vik{{v_i}_k}

\def\mjk{{m_j}_k}
\def\vjk{{v_j}_k}
\def\alphajk{{\alpha_j}_k}
\def\sigmajk{{\sigma_j^2}_k}
\def\Sigmajk{{\Sigmab_j}_k}
\def\alphajk{{\alpha_j}_k}
\def\betajk{{\beta_j}_k}

\def\mujk{{\mu_j}_k}
\def\vjk{{v_j}_k}
\def\njk{{n_j}_k}

\def\alphae{\alpha^{\epsilon}}
\def\betae{\beta^{\epsilon}}
\def\alphaez{\alpha^{\epsilon}_{0}}
\def\betaez{\beta^{\epsilon}_{0}}
\def\alphaei{\alpha^{\epsilon}_{i}}
\def\betaei{\beta^{\epsilon}_{i}}
\def\alphaeiz{\alpha^{\epsilon}_{i0}}
\def\betaeiz{\beta^{\epsilon}_{i0}}

\def\mbz{{{\mb}_z}}
\def\Sigmabz{{{\Sigmab}_z}}
\def\diag#1{\mbox{diag}\left[#1\right]}

\def\sigmah{\widehat{\sigma}}
\def\fh{\widehat{f}}
\def\sigmaz{\sigma_z}

\def\blue#1{{\color{blue}#1}}
\def\red#1{{\color{red}#1}}
\def\green#1{{\color{green}#1}}
\def\magenta#1{{\color{magenta}#1}}
\def\cyan#1{{\color{cyan}#1}}

\def\bul{\bullet}
\def\spa{\hspace*{4.7mm}}
\def\spb{\hspace*{6.3mm}}
\def\spc{\hspace*{5.3mm}}
\def\spd{\hspace*{28mm}}
\def\tir{\leftrightarrow}

\def\bullets{$\bul\spa\bul\spa\bul\spa\bul\spa\bul\spa\bul\spa\bul\spa\bul\spa\bul\spa\bul\spa\bul$}
\def\bulletsl{$\bul\tir\bul\tir\bul\tir\bul\tir\bul\tir\bul\tir\bul\tir\bul\tir\bul\tir\bul\tir\bul$}
\def\verticals{$|\spb|\spb|\spb|\spb|\spb|\spb|\spb|\spb|\spb|\spb|$}

\def\bbullets{\blue{\bullets}}
\def\rbullets{\red{\bullets}}
\def\mbullets{\magenta{\bullets}}

\def\Sbh{\widehat{\Sb}}
\def\Zbh{\widehat{\Zb}}
\def\Thetabh{\widehat{\Thetab}}

\def\bm#1{\mbox{\boldmath $#1$}}
\def\zerob{{\bm 0}}
\def\oneb{{\bm 1}}
\def\intg{\int\kern-1.1em\int}
\def\expf#1{\exp\left[ {#1} \right]}

\def\gbu{\underline{\gb}}
\def\fbu{\underline{\fb}}
\def\zbu{\underline{\zb}}
\def\epsilonbu{\underline{\epsilonb}}
\def\thetabu{\underline{\thetab}}

\def\sigmae{{\sigma_{\epsilon}}}
\def\Sigmae{{\Sigmab_{\epsilonb}}}
\def\Sigmaf{{\Sigmab_{\fb}}}
\def\Sigmag{{\Sigmab_{\gb}}}

\def\fbuh{\widehat{\fbu}}
\def\zbuh{\widehat{\fbu}}
\def\thetabuh{\widehat{\thetabu}}

\def\fbh{\widehat{\fb}}
\def\Sigmabh{\widehat{\Sigmab}}
\def\Abh{\widehat{\Ab}}
\def\thetabh{\widehat{\thetab}}

\def\Rjk{{R_j}_k}
\def\fbik{{\fb_i}_k}
\def\fbjk{{\fb_j}_k}

\def\mik{{m_i}_k}
\def\sigmaik{{\sigma_i^2}_k}
\def\Sigmaik{{\Sigmab_i}_k}
\def\alphaik{{\alpha_i}_k}
\def\betaik{{\beta_i}_k}

\def\alphaw{\alpha_\omega}
\def\lambdaw{\lambda_\omega}

\def\muik{{\mu_i}_k}
\def\vik{{v_i}_k}

\def\mjk{{m_j}_k}
\def\sigmajk{{\sigma_j^2}_k}
\def\Sigmajk{{\Sigmab_j}_k}
\def\alphajk{{\alpha_j}_k}
\def\betajk{{\beta_j}_k}

\def\mujk{{\mu_j}_k}
\def\vjk{{v_j}_k}
\def\njk{{n_j}_k}

\def\alphae{\alpha^{\epsilon}}
\def\betae{\beta^{\epsilon}}
\def\alphaez{\alpha^{\epsilon}_{0}}
\def\betaez{\beta^{\epsilon}_{0}}
\def\alphaei{\alpha^{\epsilon}_{i}}
\def\betaei{\beta^{\epsilon}_{i}}
\def\alphaeiz{\alpha^{\epsilon}_{i0}}
\def\betaeiz{\beta^{\epsilon}_{i0}}
\def\alphaeiz{\alpha^{\epsilon}_{i0}}
\def\betaeiz{\beta^{\epsilon}_{i0}}
\def\alphaeiz{\alpha^{\epsilon}_{i0}}
\def\betaeiz{\beta^{\epsilon}_{i0}}
\def\betaj{\beta_{j}}
\def\mbz{{{\mb}_z}}
\def\Sigmabz{{{\Sigmab}_z}}
\def\diag#1{\mbox{diag}\left[#1\right]}

\def\vect{\mbox{vect}}
\def\lra{\longrightarrow}

\def\gir{g_i(\rb)}
\def\fjr{f_j(\rb)}
\def\zjr{z_j(\rb)}
\def\epsilonir{\epsilon_i(\rb)}
\def\gbr{\gb(\rb)}
\def\fbr{\fb(\rb)}
\def\zbr{\zb(\rb)}
\def\epsilonbr{\epsilonb(\rb)}
\def\fbhr{\fbh(\rb)}
\def\Sigmabhr{\Sigmabh(\rb)}

\def\mbzr{\mb_{z(\rb)}}
\def\Sigmabzr{\Sigmab_{z(\rb)}}
\def\Sigmagbzr{{\Sigmab_g}_{z(\rb)}}

\def\mjzjr#1{{m_{#1}}_{z_{#1}(\rb)}}
\def\sigmajzjr#1{{\sigma_{#1}}_{z_{#1}(\rb)}}

\def\Beta{B}
\def\Betab{\bm{\Beta}}
\def\Betabe{\bm{\Beta}_{\epsilonb}}

\def\fh{\widehat{f}}
\def\sigmah{\widehat{\sigma}}
\def\sigmaei{\sigma_{\epsilon_i}^2}

\def\Xwr{X(\omega,\rb)}
\def\xwr{x(\omega,\rb)}

\def\muzw{\mu_{z}(\omega)}
\def\muzwr{\mu_{z(\rb)}(\omega)}
\def\szw{\sigma_{z}^2(\omega)}
\def\szwr{\sigma_{z(\rb)}^2(\omega)}
\def\pzw{p_{z}(\omega)}
\def\pzwr{p_{z(\rb)}(\omega)}
\def\pwr{p_{\omega}(\rb)}

\def\hszw{\widehat{\sigma_{z}^2}(\omega)}
\def\hmuzw{\widehat{\mu}_{z}(\omega)}
\def\hpzw{\widehat{p}_{z}(\omega)}

\def\muzwrl{\mu_{z(\rb)}(\omega-l)}

\def\pxwrcz{p\left(\xwr ~|~ z\right)}
\def\pxwr{p\left(\xwr\right)}

\def\Xbw{\Xb(\omega)}
\def\xbw{\xb(\omega)}
\def\uXb{\underline{\Xb}}
\def\uxb{\underline{\xb}}
\def\pxbw{p(\xbw)}
\def\puxb{p(\uxb)}
\def\pzb{P(\zb)}

\def\Szw{S_z(\omega)}
\def\mubw{\mub(\omega)}
\def\mubr{\mub(\rb)}
\def\Srw{S_{\rb}(\omega)}
\def\Swr{S_{\omega}(\rb)}
\def\Sbw{\Sb(\omega)}
\def\swr{s_{\omega}(\rb)}
\def\lwr{l_{\omega}(\rb)}
\def\sbw{\sb(\omega)}
\def\Sbr{\Sb(\rb)}
\def\mubw{\mub(\omega)}
\def\cov#1{\mbox{cov}[#1]}

\def\pdf{\emph{pdf}}
\def\pmf{\emph{pmf}}
\def\dpdx#1#2{\frac{\partial #1}{\partial #2}}

\def\REM#1{}

\def\XXwr{X(\omega,\rb)}
\def\xxwr{x(\omega,\rb)}

\def\Xwr{X_{\omega}(\rb)}
\def\xwr{x_{\omega}(\rb)}
\def\Xrw{X_{\rb}(\omega)}
\def\xrw{x_{\rb}(\omega)}

\def\Ywr{Y_{\omega}(\rb)}
\def\ywr{y_{\omega}(\rb)}
\def\Yrw{Y_{\rb}(\omega)}
\def\yrw{y_{\rb}(\omega)}

\def\Ewr{E_{\omega}(\rb)}
\def\ewr{\epsilon_{\omega}(\rb)}
\def\Erw{E_{\rb}(\omega)}
\def\erw{\epsilon_{\rb}(\omega)}

\def\Xbr{\Xb(\rb)}
\def\xbr{\xb(\rb)}
\def\xbrp{\xb'(\rb)}
\def\Xbw{\Xb(\omega)}
\def\xbw{\xb(\omega)}
\def\Sbr{\Sb(\rb)}
\def\Bbr{\Bb( \rb)}
\def\sbr{\sb(\rb)}
\def\Sbw{\Sb(\omega)}
\def\sbw{\sb(\omega)}
\def\usw{\underline{s}_\omega}
\def\usb{\underline{\sb}}
\def\ueb{\underline{\epsilonb}}
\def\uXb{\underline{\Xb}}
\def\uxb{\underline{\xb}}
\def\uXbz{\underline{\Xb}_z}
\def\uxbz{\underline{\xb}_z}
\def\ucb{\underline{\cb}}
\def\uSb{\underline{\Sb}}
\def\ulb{\underline{\lb}}
\def\lbr{\lb(\rb)}
\def\uthetab{\underline{\thetab}}
\def\uthetabz{\underline{\thetab}_z}
\def\zwr{z_{\omega}(\rb)}
\def\Ywr{Y_{\omega}(\rb)}
\def\ywr{y_{\omega}(\rb)}
\def\Yrw{Y_{\rb}(\omega)}
\def\yrw{y_{\rb}(\omega)}
\def\Ybr{\Yb(\rb)}
\def\ybr{\yb(\rb)}
\def\Ybw{\Yb(\omega)}
\def\ybw{\yb(\omega)}
\def\uYb{\underline{\Yb}}
\def\uyb{\underline{\yb}}
\def\uYbz{\underline{\Yb}_z}
\def\uybz{\underline{\yb}_z}

\def\Ewr{E_{\omega}(\rb)}
\def\ewr{\epsilon_{\omega}(\rb)}
\def\Erw{E_{\rb}(\omega)}
\def\erw{\epsilon_{\rb}(\omega)}
\def\Ebr{Eb(\rb)}
\def\ebr{\epsilonb(\rb)}
\def\Ebw{Eb(\omega)}
\def\ebw{\epsilonb(\omega)}
\def\uEb{\underline{\Eb}}

\def\Erwp{E_{\rb}(\omega')}
\def\Xrwp{X_{\rb}(\omega')}

\def\xwmr{x_{\omega-1}(\rb)}
\def\muzwm{\mu_{z}(\omega-1)}

\def\rhoz{\rho_z}
\def\sez{\sigma_{\eta_z}^2}

\def\xwrp{x_{\omega}(\rb')}
\def\xrwp{x_{\rb}(\omega')}
\def\xwpr{x_{\omega'}(\rb)}
\def\xwprp{x_{\omega'}(\rb')}
\def\Xwrp{x_{\omega}(\rb')}

\def\Xwrp{X_{\omega}(\rb')}
\def\Xwpr{X_{\omega'}(\rb)}
\def\Xwprp{X_{\omega'}(\rb')}

\def\lwrrp{l_{\omega}(\rb,\rb')}
\def\swrp{s_{\omega}(\rb')}
\def\mwz{m_{\omega z}}
\def\sigmawz{\sigma_{\omega z}^2}
\def\Zwr{Z_{\omega}(\rb)}
\def\uzb{\underline{\zb}}

\def\alphaw{\alpha_{\omega}}
\def\lambdaw{\lambda_{\omega}}
\def\Lambdarrp{\Lambda_{\rb}(\rb')}

\def\usw{\underline{s}_\omega}

\def\uqb{\underline{\qb}}
\def\ucb{\underline{\cb}}

\def\cbr{\cb(\rb)}

\def\barsb{\bar{\sb}}
\def\barcb{\bar{\cb}}
\def\barlb{\bar{\lb}}
\def\barqb{\bar{\qb}}
\def\barzb{\bar{\zb}}

\def\barsbr{\barsb(\rb)}
\def\barsbrp{\barsb(\rb')}
\def\barcbr{\barcb(\rb)}
\def\barlbr{\barlb(\rb)}
\def\barqbr{\barqb(\rb)}
\def\barzbr{\barzb(\rb)}

\def\barlbrrp{\barlb(\rb,\rb')}
\def\barqbrrp{\barqb(\rb,\rb')}

\def\barswr{\bar{s}_\omega(\rb)}
\def\barlwr{\bar{l}_\omega(\rb)}
\def\barqwr{\bar{q}_\omega(\rb)}
\def\barzwr{\bar{z}_\omega(\rb)}

\def\barswrp{\bar{s}_\omega(\rb')}
\def\barlwrp{\bar{l}_\omega(\rb')}
\def\barqwrp{\bar{q}_\omega(\rb')}
\def\barzwrp{\bar{z}_\omega(\rb')}
\def\zbr{\zb(\rb)}
\def\zbrp{\zb(\rb')}
\def\zrb{z(\rb)}
\def\zrpp{z(\rb')}

\def\sbrp{\sb(\rb')}
\def\srb{s(\rb)}
\def\srpp{s(\rb')}

\def\barzbrp{\bar{\zb}(\rb')}
\def\barzrb{\bar{z}(\rb)}
\def\barzrpp{\bar{z}(\rb')}
\def\srb{s(\rb)}
\def\barsrb{\bar{s}(\rb)}
\def\barsrbp{\bar{s}(\rb')}

\def\blue#1{#1}
\def\magenta#1{#1}

\def\bul{\bullet}
\def\spa{\hspace*{2.4mm}}
\def\spb{\hspace*{4.0mm}}
\def\spc{\hspace*{3.1mm}}
\def\tir{\hspace*{-1.2mm}\leftrightarrow\hspace*{-1.1mm}}

\def\bullets{$\bul\spa\bul\spa\bul\spa\bul\spa\bul\spa\bul\spa\bul\spa\bul\spa\bul\spa\bul\spa\bul$}
\def\bulletsl{$\bul\tir\bul\tir\bul\tir\bul\tir\bul\tir\bul\tir\bul\tir\bul\tir\bul\tir\bul\tir\bul$}
\def\verticals{$|\spb|\spb|\spb|\spb|\spb|\spb|\spb|\spb|\spb|\spb|$}
\def\undeuxtrois{1\spc1\spc1\spc1 \spa2\spc2 \spa3\spc3\spc3 \spc1\spc1}

\def\bbullets{\blue{\bullets}}
\def\rbullets{\red{\bullets}}
\def\mbullets{\magenta{\bullets}}

\def\Sbh{\widehat{\Sb}}
\def\Zbh{\widehat{\Zb}}
\def\Thetabh{\widehat{\Thetab}}
\def\Sigmabe{\Sigmab_{\epsilon}}
\def\zbh{\widehat{\zb}}
\def\usbh{\widehat{\usb}}
\def\uthetabh{\widehat{\uthetab}}

\def\mbkr{\mb_k(\rb)}
\def\Sigmabkr{\Sigmab_k(\rb)}

\title{Hierarchical Markovian models for hyperspectral image segmentation}
\name{Ali MOHAMMAD-DJAFARI, Adel MOHAMMADPOOR and Nadia BALI}
\address{Laboratoire des Signaux et Syst\`emes, \\ 
Unit\'e mixte de recherche 8506 (CNRS-Sup\'elec-UPS) \\  
Sup\'elec, Plateau de Moulon, 3 rue Joliot Curie, 91192 Gif-sur-Yvette, 
France.}

\begin{document}
\maketitle

\begin{abstract}
Hyperspectral images can be represented either as a set of images or as a set of spectra. Spectral classification and segmentation and data reduction are the main problems in hyperspectral image analysis. In this paper we propose a Bayesian estimation approach with an appropriate hiearchical model with hidden markovian variables which gives the possibility to jointly do data reduction, spectral classification and image segmentation. 
In the proposed model, the desired independent components are piecewise homogeneous images which share the same common hidden segmentation variable. Thus, the joint Bayesian estimation of this hidden variable as well as the sources and the mixing matrix of the source separation problem gives a solution for all the three problems of dimensionality reduction, spectra classification and segmentation of hyperspectral images. 
A few simulation results illustrate the performances of the proposed method compared to other classical methods usually used in hyperspectral image processing. 
\end{abstract}

\section{Introduction}
Hyperspectral images data can be represented either as a set of images 
$x_{\omega}(\rb)$ or as a set of spectra $x_{\rb}(\omega)$ 
where $\omega\in\Omega$ indexes the wavelength and $\rb\in\Rc$ a pixel position 
\cite{Sasaki-1987,Parra-2000,Bali05}. 
In both representations, the data are dependent in both spatial positions and in spectral wavelength variable. 
Classical methods of hyperspectral image analysis try either to classify the spectra $x_{\omega}(\rb)$ 
in $K$ classes $\{a_k(\omega), k=1,\cdots,K\}$ or to classify the images $x_{\omega}(\rb)$ 
in $K$ classes $\{s_k(\rb), j=1,\cdots,N\}$, using the classical classification methods 
such as distance based methods (like $K$-means) or probabilistic methods using the mixture of Gaussian (MoG) 
modeling of the data. 
These methods thus either neglect the spatial structure of the spectra or the spectral natures 
of the pixels along the wavelength bands.  

The dimensionality reduction problem in hyperspectral images can be written as:
\beq \label{EQ1}
 x_{\rb}(\omega)=\sum_{k=1}^K  s_k(\rb) \; a_k(\omega) +  \epsilon_{\rb}(\omega),
\eeq
where the $a_k(\omega)$ are the $K$ spectral source components and $s_k(\rb)$ are their associated images. 

This relation, when discretized, can be written as follows:
\beq \label{EQ2}
 \xbr=\Ab\sbr+\ebr
\eeq
$\xbr=\{x_{i}(\rb), i=1,\cdots,M\}$ is the set of $M$ observed images in different bands $\omega_i$, 
$\Ab$ is the mixing matrix of dimensions $(M,K)$ whose columns are composed of the spectra $a_k(\omega)$, 
$\sbr=\{s_{k}(\rb), k=1,\cdots,K\}$ is the set of $K$ unknown components (source images) 
and $\ebr=\{\epsilon_{i}(\rb), i=1,\cdots,m\}$ represents the errors. 

The main objective in unsupervised classification of the spectra is to find both the spectra 
 $a_k(\omega)$ and their associated image components $s_{k}(\rb)$. This problem, written as in equation 
(\ref{EQ2}) is recognized as the Blind Source Separation (BSS) in signal processing community, for which, 
many general solutions such as Principal Components Analysis (PCA) and Independent Components Analysis (ICA) 
have been proposed. 
However these general purpose methods do not account for the specificity of the hyperspectral images. 

Indeed, as we mentioned, neither the classical methods of spectra or images classification nor the 
PCA and ICA methods of BSS give satisfactory results for hyperspectral images. The reasons are that, 
in the first category of methods either they account for spatial or for spectral properties and not for 
both of them simultaneously, and PCA and ICA methods do not account for the specificity of the mixing matrix and the sources. 

In this paper, we propose to use this specificity of the hyperspectral images and 
consider the dimensionality reduction problem as the blind sources separation (BSS) of equation \ref{EQ2} and use  
a Bayesian estimation framework with a hierarchical model for the sources with a 
common hidden classification variable which is modelled as a Potts-Markov field. 
The joint estimation of this hidden variable, the sources and the mixing matrix of the BSS problem 
gives a solution for all the three problems of dimensionality reduction, spectra classification and segmentation of 
hyperspectral images. 

\section{Proposed model and method}
We propose to consider the equation (\ref{EQ2}) written in the following vector form: 
\beq  \label{EQ3}
 \uxb=\Ab\usb+\epsilonb
\eeq
where we used $\uxb=\{\xbr, \rb\in\Rc\}$, $\usb=\{\sbr, \rb\in\Rc\}$ and 
$\ueb=\{\ebr, \rb\in\Rc\}$ and we are going to account for the specificity of the hyperspectral images 
through a probabilistic modeling of all the unknowns, starting by assuming that the errors $\ebr$ are centered, white, Gaussian with covariance matrix 
$\Sigmabe=\diag{\sigma_{\epsilon_1}^2,..,\sigma_{\epsilon_M}^2}$. 
This leads to 
\beq  \label{EQ4}
p(\uxb|\usb,\Ab,\Sigmabe)=\prod_{\rb} \Nc(\Ab\sbr,\Sigmabe)
\eeq
The next step is to model the sources. 
As we mentioned in the introduction, we want to impose to all these sources $\sbr$ to be piecewise homogeneous 
and share the same common segmentation, where the pixels in each region are considered to be homogeneous and associated to 
a particular spectrum representing the type of the material in that region. We also want 
that those spectra be classified in $K$ distinct classes, thus all the pixels in regions associated with a particular spectrum 
share some common statistical parameters. 
This can be achieved through the introduction of a discrete valued hidden variable $z(\rb)$ representing the labels 
associated to each type of material and thus assuming the following:
\beq  \label{EQ5}
p(s_j(\rb)| z(\rb)=k))=\Nc(\mjk,\sigmajk), \quad k=1,\cdots,K
\eeq
with the following Potts-Markov field model
\beq  \label{EQ6}
p(\zb)\propto
\expf{\beta \sum_{\rb}\sum_{\rb'\in\Vc(\rb)}\delta(z(\rb)-z(\rb'))}
\eeq
where $\zb=\{z(\rb),\rb\in\Rc\}$ represents the common segmentation of the sources and the data. 
The parameter $\beta$ controls the mean size of those regions. 

We may note that, assuming \emph{a priori} that the sources are mutually independent 
and that pixels in each class $k$ are independent form those of class $k'$, 
we have
\beq 
p(\usb|\zb)= \sum_k \sum_{\rb\in\Rc_k} \sum_j p(s_j(\rb)|z(\rb)=k))  
\eeq
where $\Rc_k=\{\rb : z(\rb)=k\}$ and $\Rc=\cup_k \Rc_k$.

To insure that each image $s_j(\rb)$ is only non-zero in those regions associated with the $k$th spectrum, we impose $K=n$ and $\mjk=0, \forall j\not=k$ and $\sigmajk=0, \forall j\not=k$. We may then write 
\beq  \label{EQ7}
p(\usb|\zb)= \sum_{\rb} p(\sbr|z(\rb)=k))= \sum_{\rb} \Nc(\mbkr,\Sigmabkr)
\eeq
where $\mbkr$ is a vector of size $n$ with all elements equal to zero except the $k$-th element $k=z(\rb)$ and $\Sigmabkr$ is a diagonal matrix of size $n\times n$ with all elements equal to zero except the $k$-th main diagonal element where $k=z(\rb)$. 

Combining the observed data model (\ref{EQ3}) and the sources model (\ref{EQ6}) of the previous section, 
we obtain the following hierarchical model:

\bfig[htb]
\bcc
\btabu{@{}l@{~}l@{}}
\bullets            & $x_i(\rb)|\sb(\rb)$
\\[-6pt] \verticals & \\[-6pt] 
\blue{\bullets}     & \blue{$s_j(\rb)|z(\rb)$} 
\\[-6pt] \verticals & \\[-6pt] 
\magenta{\bulletsl} & \magenta{$z_j(\rb)|z_j(\rb'),\rb'\in\Vc(\rb)$}
\\[-1pt]
\magenta{\undeuxtrois} & \magenta{$z(\rb)=\{1,\cdots,K\}$} 
\\[-12pt]
\etabu
\ecc
\caption{Proposed hierarchical model for hyperspectral images: the sources $s_j(\rb)$ are hidden variables for the data  $x_i(\rb)$ and the common classification and segmentation variables $z(\rb)$ is a hidden variable for the sources.} 
\efig

\vspace*{-12pt}
\section{Bayesian estimation framework}
Using the prior data model (\ref{EQ5}), the prior source model (\ref{EQ6}) and the prior Potts-Markov model 
(\ref{EQ7}) and also assigning appropriate prior probability laws $p(\Ab)$ and $p(\uthetab)$ to the hyperparameters  $\uthetab=\{\thetab_{\epsilon},\thetab_s\}$ 
where $\thetab_{\epsilon}=\Rbe$ and $\thetab_s=\{(\mjk,\sigmajk)\}$, we obtain an expression for the 
posterior law 
\beq
p(\usb,\zb,\Ab,\uthetab|\uxb)\propto p(\uxb|\usb,\Ab,\thetab_{\epsilon})\,p(\usb|\zb,\thetab_s)\,p(\zb)\,p(\Ab)\,p(\uthetab) 
\eeq 
I this paper, we used conjugate priors for all of them, 
i.e., Gaussian for the elements of $\Ab$, Gaussian for the means $\mjk$ and inverse Gamma 
for the variances $\sigmajk$ as well as for the noise variances $\sigmae_i$.  

When given the expression of the posterior law, we can then use it to define an estimator such as 
Joint Maximum A Posteriori (JMAP) 
or the Posterior Means (PM) for all the unknowns. 
The first needs optimization algorithms and the second integration methods. 
Both are computationally demanding. Alternate optimization is generally used for the first while the 
MCMC techniques are used for the second.  

In this work, we propose to separate the unknowns in two sets $(\usb,\zb)$ and $(\Ab,\uthetab)$ and then 
use the following iterative algorithm:
\bit
\item Estimate $(\usb,\zb)$ using $p(\usb,\zb|\Abh,\uthetabh,\uxb)$ by 
\\ 
\(
\barr{lll}
\usbh \sim p(\usb|\zbh,\Abh,\uthetabh,\uxb) & \mbox{and} &
\zbh  \sim p(\zb|\Abh,\uthetabh,\uxb) 
\earr
\)
\item Estimate $(\Ab,\uthetab)$ using $p(\Ab,\uthetab|\usbh,\zbh,\uxb)$ by 
\\ 
\(
\barr{lll}
\Abh     \sim p(\Ab|\usbh,\zbh,\thetabh,\uxb) & \mbox{and} &
\uthetabh\sim p(\uthetab|\usbh,\zbh,\Abh,\uxb) 
\earr
\)
\eit 
In this algorithm, $\sim$ represents either $argmax$ or \emph{generate sample using} or still 
\emph{compute the Mean Field Approximation (MFA)}. 
To implement this algorithm, we need the following expressions:

\smallskip\noindent$\bullet$\quad  
\(
p(\usb|\zb,\Ab,\uthetab,\uxb)\propto p(\uxb|\usb,\Ab,\Sigmabe) \;  p(\usb|\zb,\uthetab)
\). \\ 
It is then easy to see that $p(\usb|\zb,\Ab,\uthetab,\uxb)$ is separable in $\rb$:
\beqn
p(\usb|\zb,\uthetab,\uxb)
&=&\prod_{\rb} p(\sbr|z(\rb),\thetab,\xbr) \nonumber \\
&=&\prod_{\rb} \Nc(\barsbr,\Bbr)
\eeqn
with \\[-12pt]
\beq \label{EQ9}
\left\{\barr{l}
\Bbr=\left[\Ab^t\Sigmabe^{-1}\Ab + \Sigmabzr^{-1}\right]^{-1}
\\
\barsbr=\Bbr[\Ab^t\Sigmabe^{-1}\xbr+\Sigmabzr^{-1}\mbzr]
\\
\earr\right.
\eeq

In this relation $\mbzr$ is a vector of size $n$ with all elements equal to zero except the $k$-th element where $k=z(\rb)$ and $\Sigmabzr$ is a diagonal matrix of size $n\times n$ with all elements equal to zero except the $k$-th diagonal where $k=z(\rb)$. 

\smallskip\noindent$\bullet$\quad 
\(p(\zb|\Ab,\uthetab,\uxb)\propto p(\uxb|\zb,\Ab,\uthetab)\;p(\zb)\), \quad where 
\beqn
p(\uxb|\uzb,\Ab,\uthetab)
&=& \prod_{\rb} p(\xbr|z(\rb),\Ab,\uthetab) \\ 
&=& \prod_{\rb} \Nc(\Ab\mbzr,\Ab\Sigmabzr\Ab^{t}+\Sigmae). \nonumber
\eeqn
It is then easy to see that, even if $p(\uxb|\uzb,\Ab,\uthetab)$ is separable in $\rb$,  $p(\zb|\Ab,\uthetab,\uxb)$ is not and it has the same markovian structure that $p(\zb)$. 

\smallskip\noindent$\bullet$\quad  
\( 
p(\Ab|\zb,\uthetab,\uxb)\propto p(\uxb|\zb,\Ab,\uthetab)\; p(\Ab).  
\)\\ 
It is easy to see that, with a Gaussian or uniform prior for $p(\Ab)$ we obtain a Gaussian expression for this posterior law. Indeed, with an uniform prior, the posterior mean is equivalent to the posterior mode and equivalent to the Maximum Likelihood (ML) estimate $\Abh=\argmax{\Ab}{p(\uxb|\zb,\Ab,\uthetab)}$ 
whose expression is:
\[
\Abh=\left[\sum_{\rb}\xbr\barsb'(\rb) \right] \left[ \sum_{\rb}\barsbr\barsb'(\rb)+\Bbr \right]^{-1}
\]
where $\barsbr$ and $\Bbr$ are given by (\ref{EQ9}).

\smallskip\noindent$\bullet$\quad  
\( 
p(\Rbe|\zb,\Ab,\uthetab,\uxb)\propto p(\uxb|\zb,\Ab,\uthetab)\; p(\Rbe). 
\)\\ 
It is also easy to show that, with an uniform prior on the logarithmic scale or an inverse gamma prior for the noise variances, the posterior is also an inverse gamma. 

\smallskip\noindent$\bullet$\quad  
$p(\uthetab|\zb,\Ab,\uxb)\propto p(\uxb|\zb,\Ab,\uthetab)\, p(\uthetab) $
\\ 
Again here, using the conjugate priors for the means $\mjk$ and inverse gamma for the variances $\sigmajk$ 
we can obtain easily the expressions of the posterior laws for them. 

Details of the expressions of $p(\Ab|\zb,\uthetab,\uxb)$, $p(\Rbe|\zb,\Ab,\uthetab,\uxb)$ and 
$p(\uthetab|\zb,\Ab,\uxb)$ as well as their modes and means can be found in \cite{Snoussi03a}.

\section{Computational considerations and Mean Field Approximation}
As we can see, the expression of the conditional posterior of the sources is separable in $\rb$ but this is not the case for the conditional posterior of the hidden variable $z(\rb)$. 
So, even if it is possible to generate samples 
from this posterior using a Gibbs sampling scheme, the cost of the computation is very high for 
real applications. 
The Mean Field Approximation (MFA) then becomes a natural tool for obtaining approximate solutions with lower computational cost. 

The mean field approximation is a general method for approximating the expectation of a Markov random variable. 
The idea consists in, when considering a pixel, to neglect the fluctuation of its neighbor pixels by fixing them 
to their mean values.
\cite{Zhang93,Landgrebe02}.
Another interpretation of the MFA is to approximate a non separable 
\beqnx
p(\zb)
&\propto& \expf{ \beta \sum_{\rb}\sum_{\rb'}\delta(z(\rb)-z(\rb'))}\\ 
&\propto& \prod_{\rb} p(z(\rb)|z(\rb'), \rb'\in \Vc(\rb))
\eeqnx
with the following separable one: 
\[
q(\zb) \propto \prod_{\rb} q(z(\rb)|\bar{z}(\rb'), \rb'\in \Vc(\rb))
\]
where $\bar{z}(\rb')$ is the expected value of $z(\rb')$ computed using $q(zb)$. 
This approximate separable expression is obtained in such a way to minimize 
$KL(p,q)$ for a given class of separable distributions $q \in \Qb$.
 
Using now this approximation in the expression of the conditional posterior law 
$p(\zb|\Ab,\uthetab,\uxb)$ gives the separable MFA 
\[
q(\zb|\Ab,\uthetab,\uxb)
=\prod_{\rb}  q(z(\rb)|\bar{z}(\rb'), \rb'\in \Vc(\rb),\Ab,\thetab,\xbr)
\]

\vspace*{-10pt}\noindent
where \quad  
$q(z(\rb)|\bar{z}(\rb'), \rb'\in \Vc(\rb),\Ab,\thetab,\xbr)=$ \\ 
\hspace*{2cm} $p(\xbr|z(\rb),\Ab,\thetab)\; q(z(\rb)|\bar{z}(\rb'), \rb'\in\Vc(\rb))$ \\ 
and $\bar{z}(\rb)$ can be computed by
\[
\bar{z}(\rb)=\frac{
\sum_{z(\rb)} z(\rb) \; q(z(\rb)|\bar{z}(\rb'), \rb'\in \Vc(\rb),\Ab,\thetab,\xbr)}
{\sum_{z(\rb)} q(z(\rb)|\bar{z}(\rb'), \rb'\in \Vc(\rb),\Ab,\thetab,\xbr)}
\]

\vspace*{-12pt}
\section{Simulation results}
The main objectives of these simulations are: 
first to show that the proposed algorithm gives the desired results, and second to compare its relative performances with respect to some classical methods. 
For this purpose, first we generated some simulated data according to the data generatin model, i.e.; 
starting by generating $z(\rb)$, then the sources $\sbr$, then using some given spectral signatures 
obtained from real materials construct the mixing matrix $\Ab$ and finally generate data $\xbr$. 
Fig.~2 shows an example of such data generated with the following parameters: $m=32, n=4, K=4$ and SNR=20~dB 
and Fig.~3 shows a comparison of the results obtained by two classical spectral and image classification methods using the classical $K$-means with the results obtained by the proposed method. 
Some other simulated results as well as the results obtained on real data will be given in near future. 

\bfig[htb]
\btabu{@{}c@{}cc@{}}
\includegraphics[width=25mm,height=25mm]{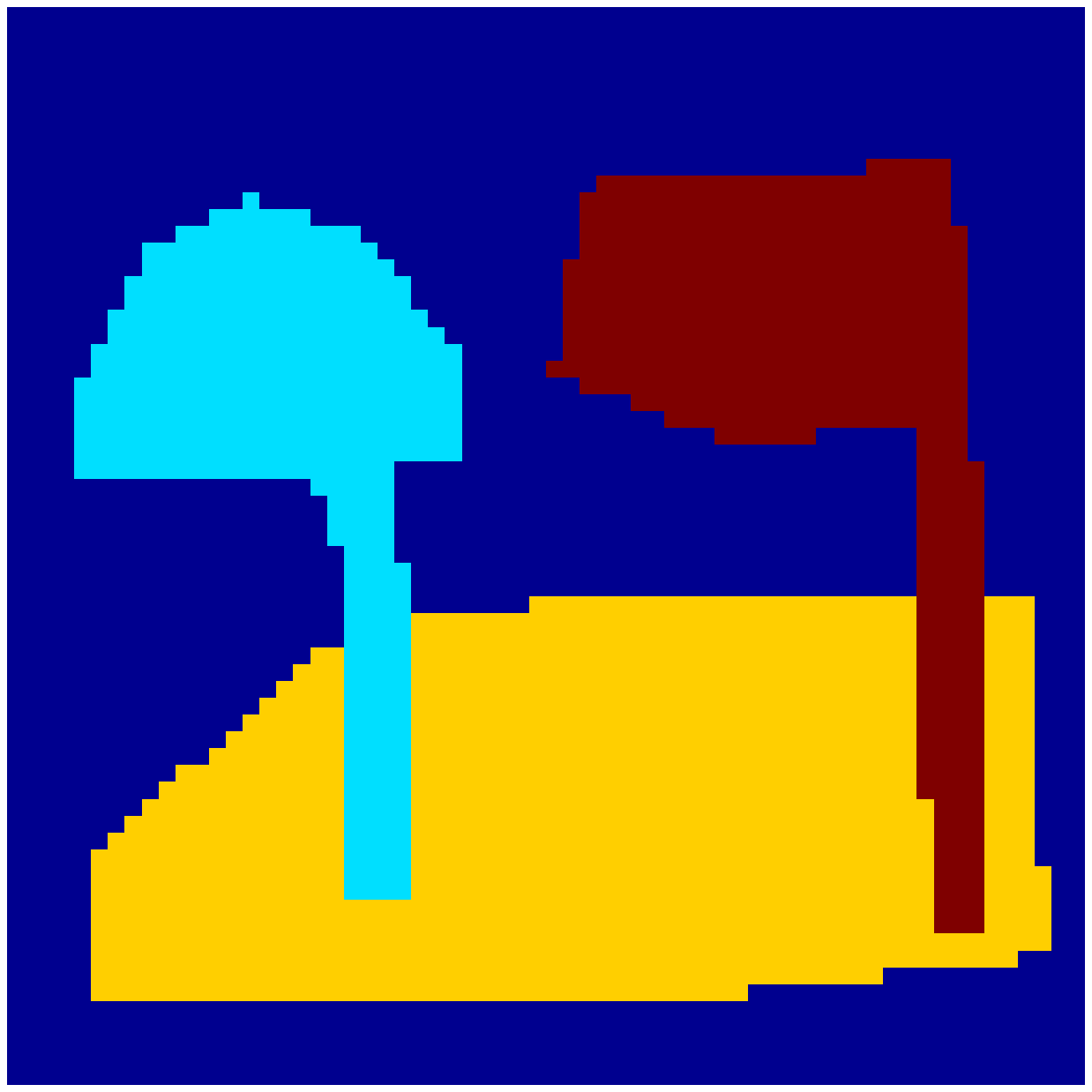} &
\includegraphics[width=25mm,height=25mm]{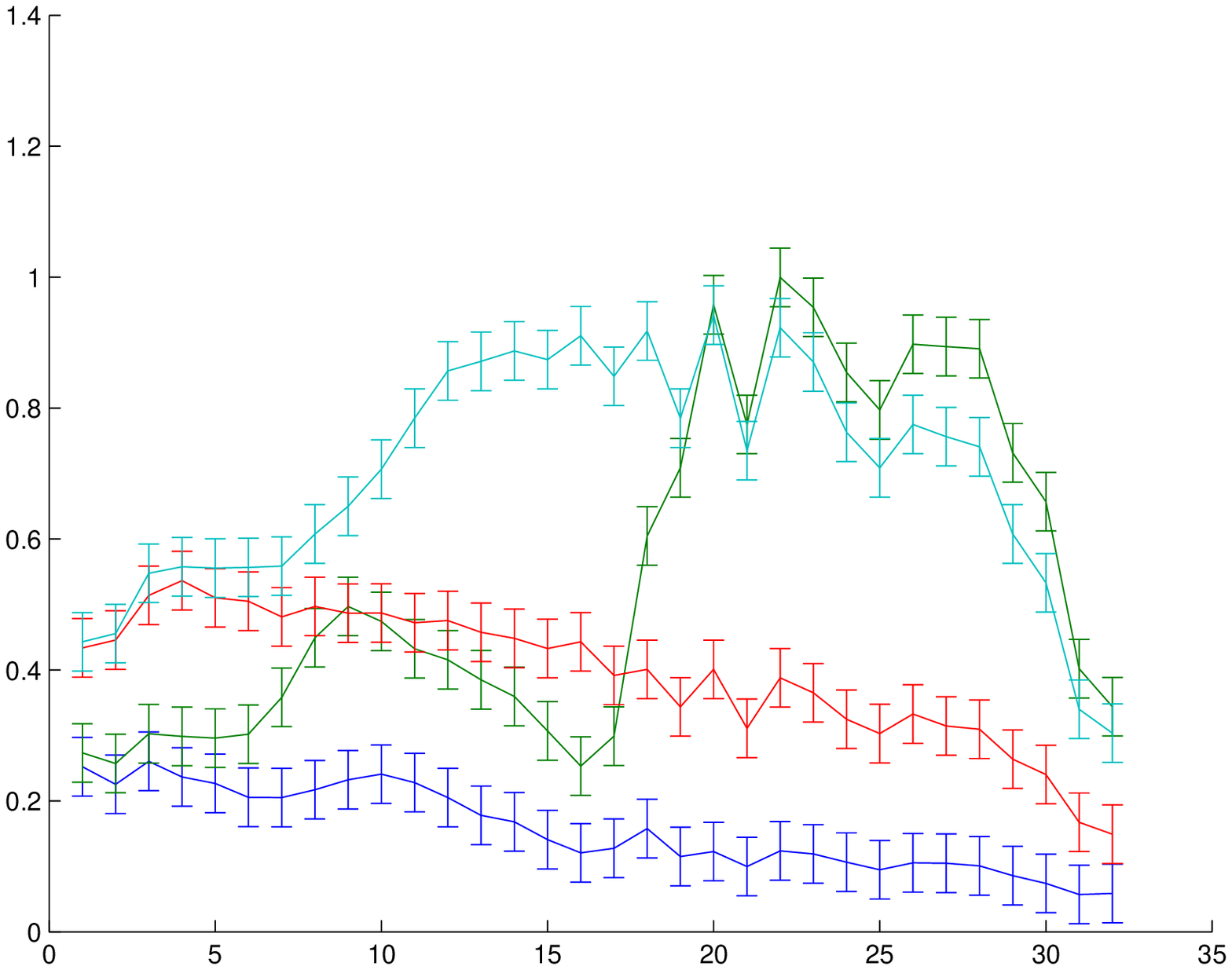} & 
\includegraphics[width=30mm,height=25mm]{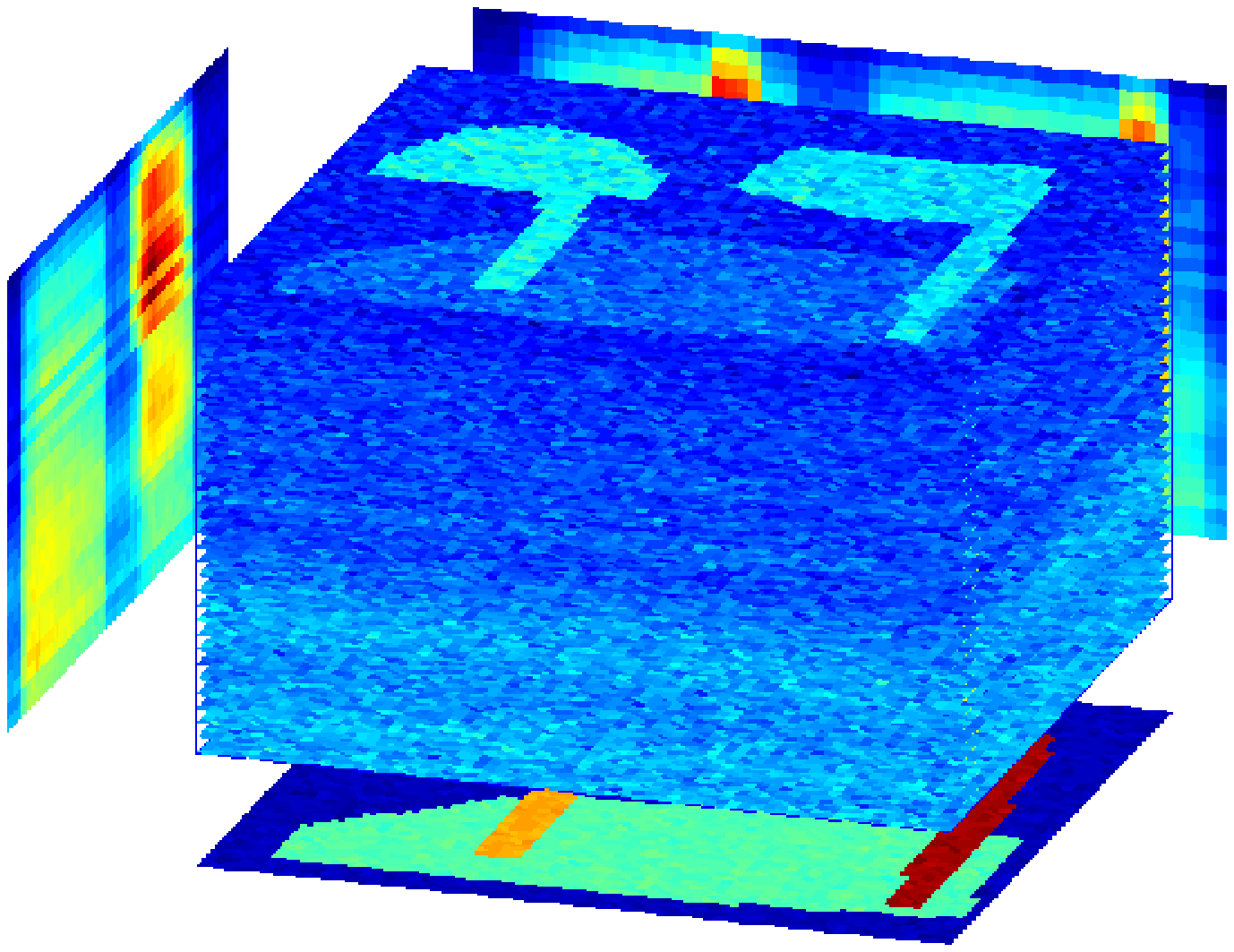}
\\  
\includegraphics[width=25mm,height=25mm]{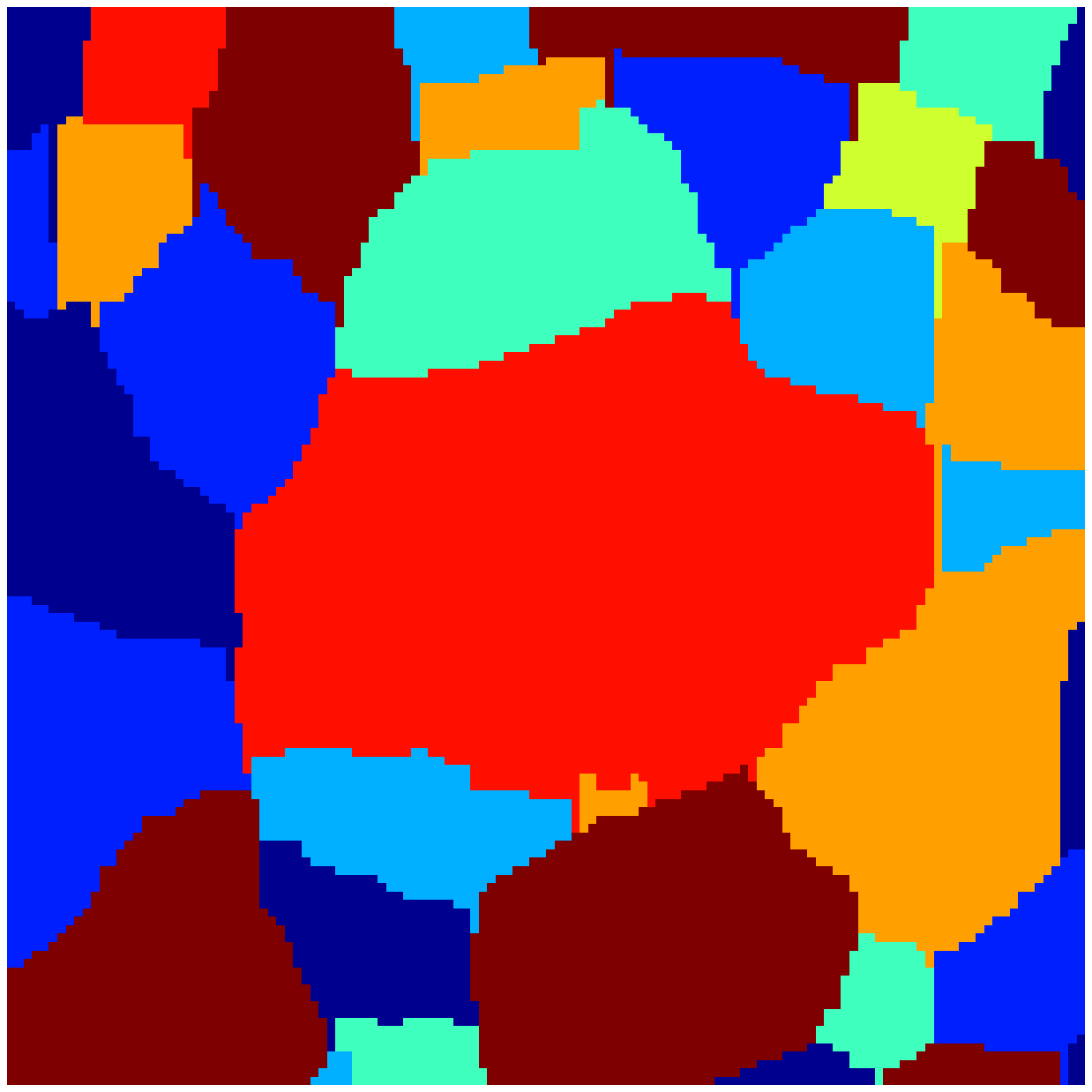} &
\includegraphics[width=25mm,height=25mm]{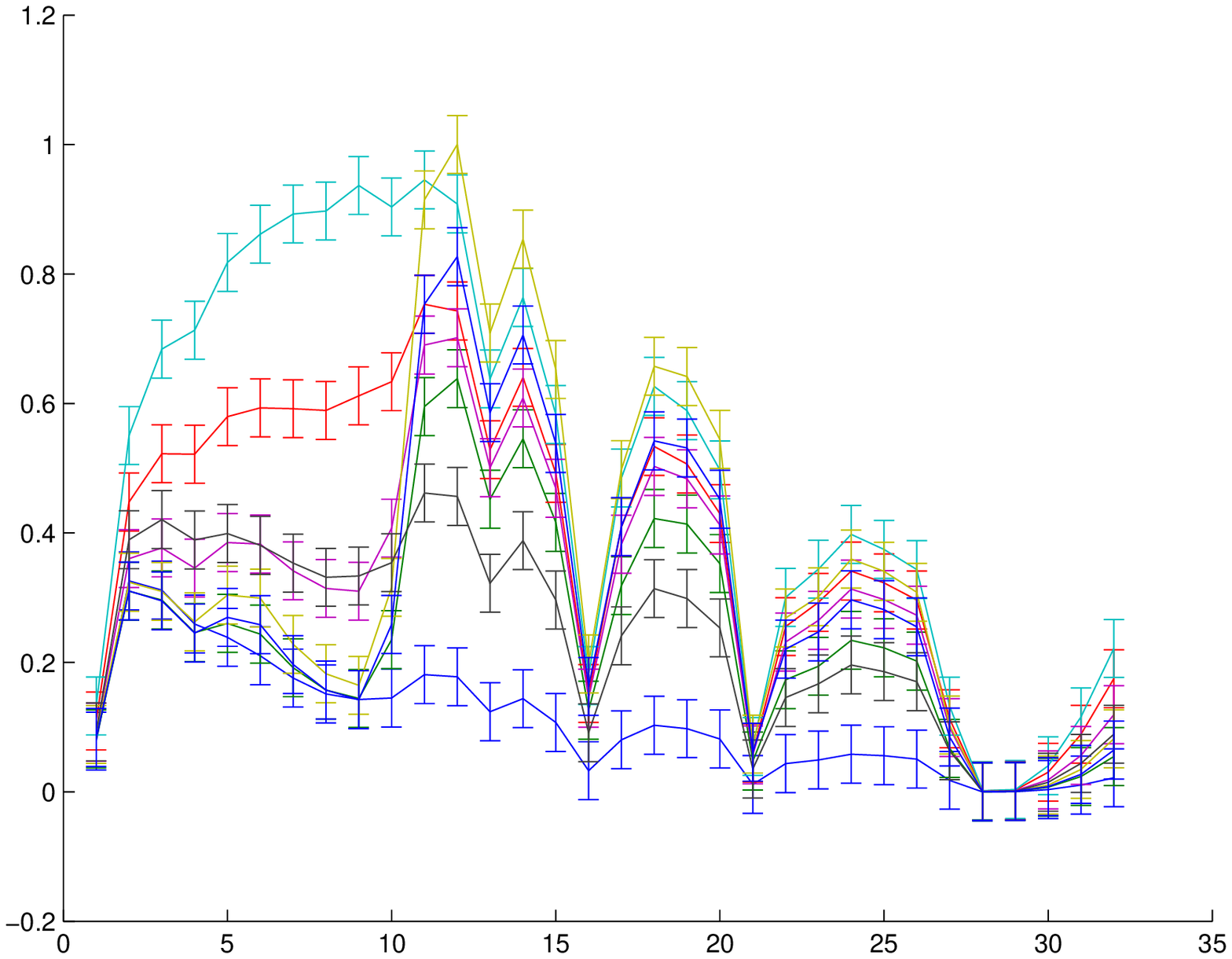} &
\includegraphics[width=30mm,height=25mm]{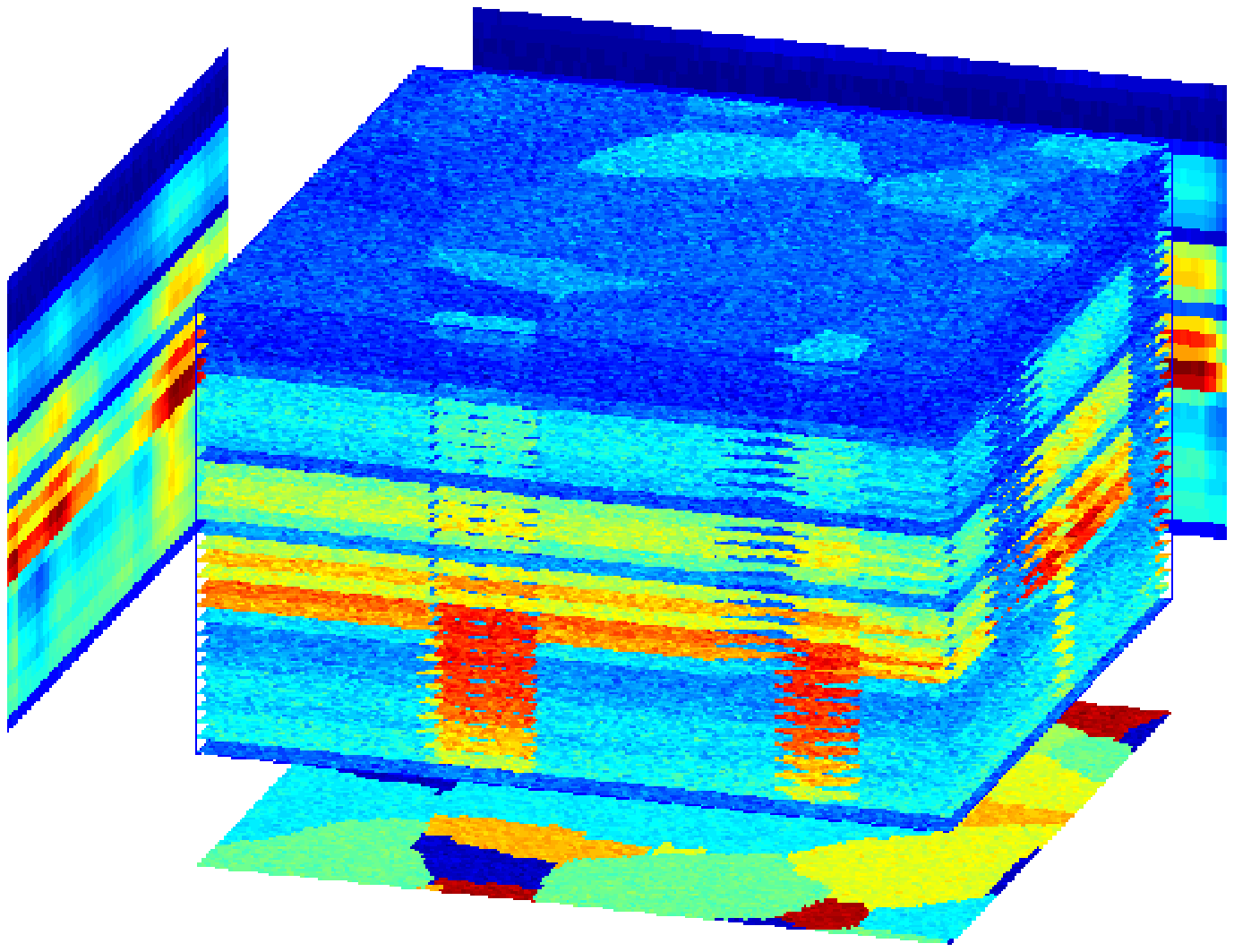} 
\\[-6pt] 
a & b & c 
\\[-12pt]
\etabu
\caption{Two examples of data generating process: 
a) $z(\rb)$ b) spectral signatures used to construct the mixing matrix $\Ab$ and c) $m=32$ images. 
Upper row: $K=4$ and image sizes (64x64). Lower row: $K=8$ and image sizes (128x128).}
\efig

\bfig[hbt]
\bcc
\btabu{@{}c@{}c@{}c@{}c@{}c@{}}
\includegraphics[width=25mm,height=25mm]{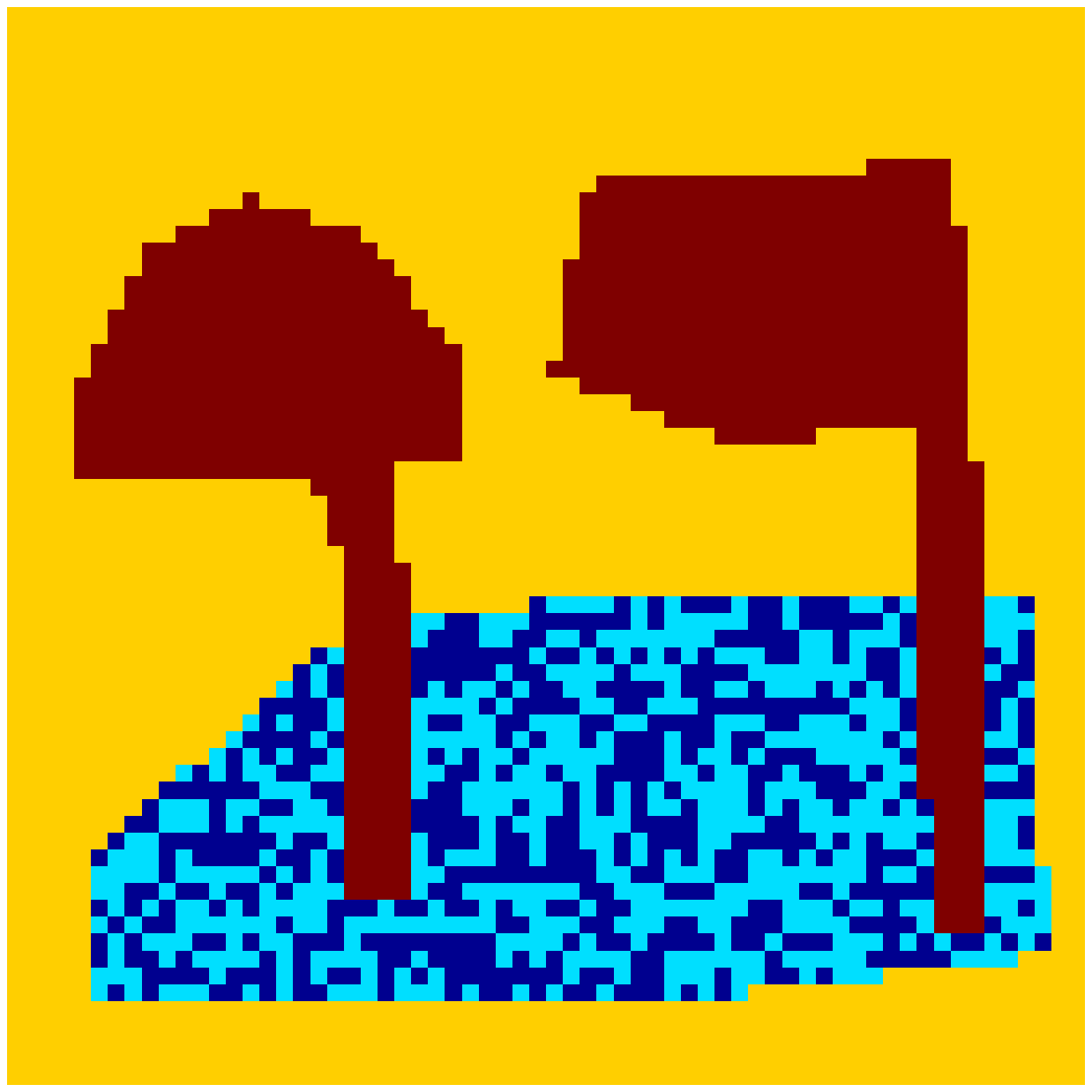}&
\includegraphics[width=25mm,height=25mm]{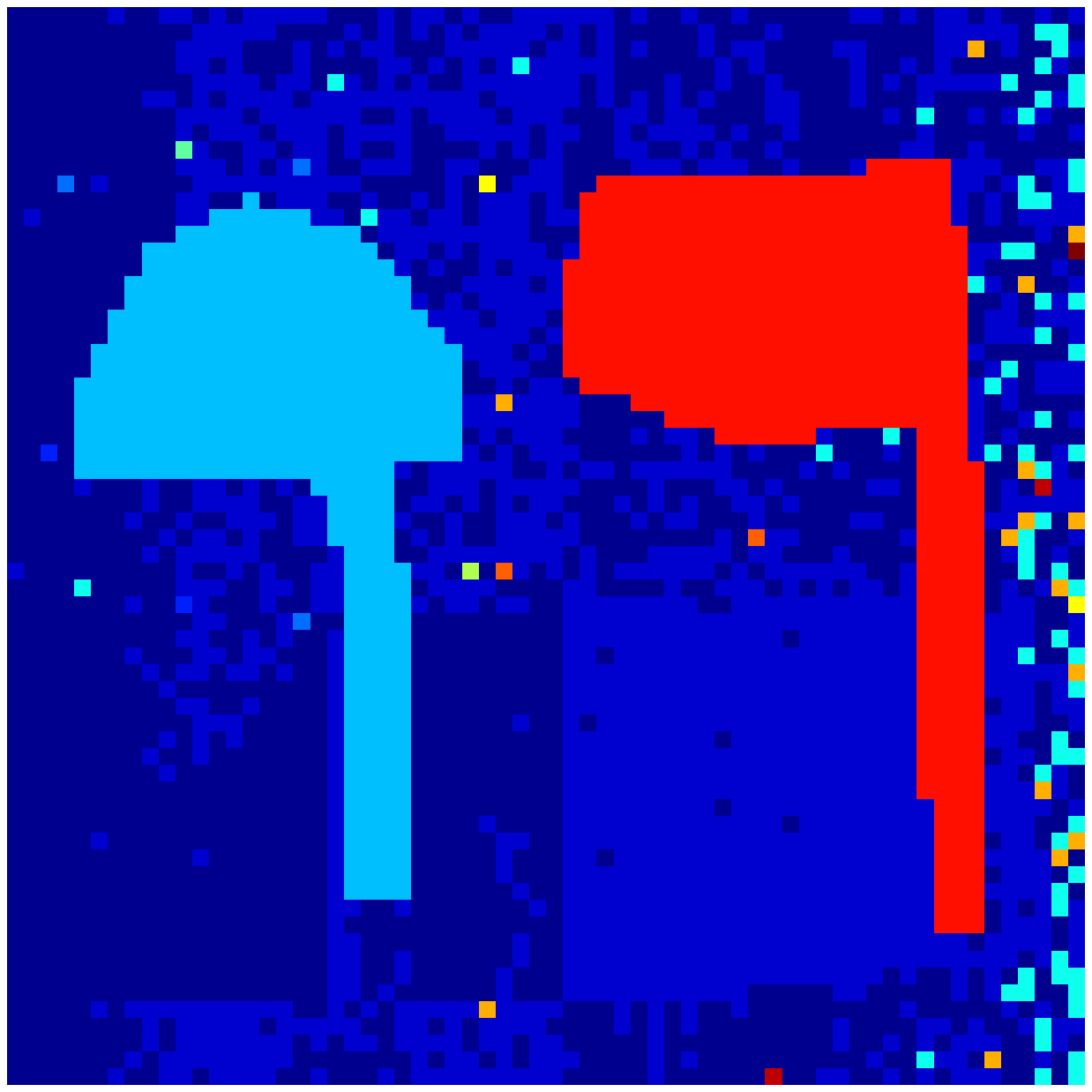}&
\includegraphics[width=25mm,height=25mm]{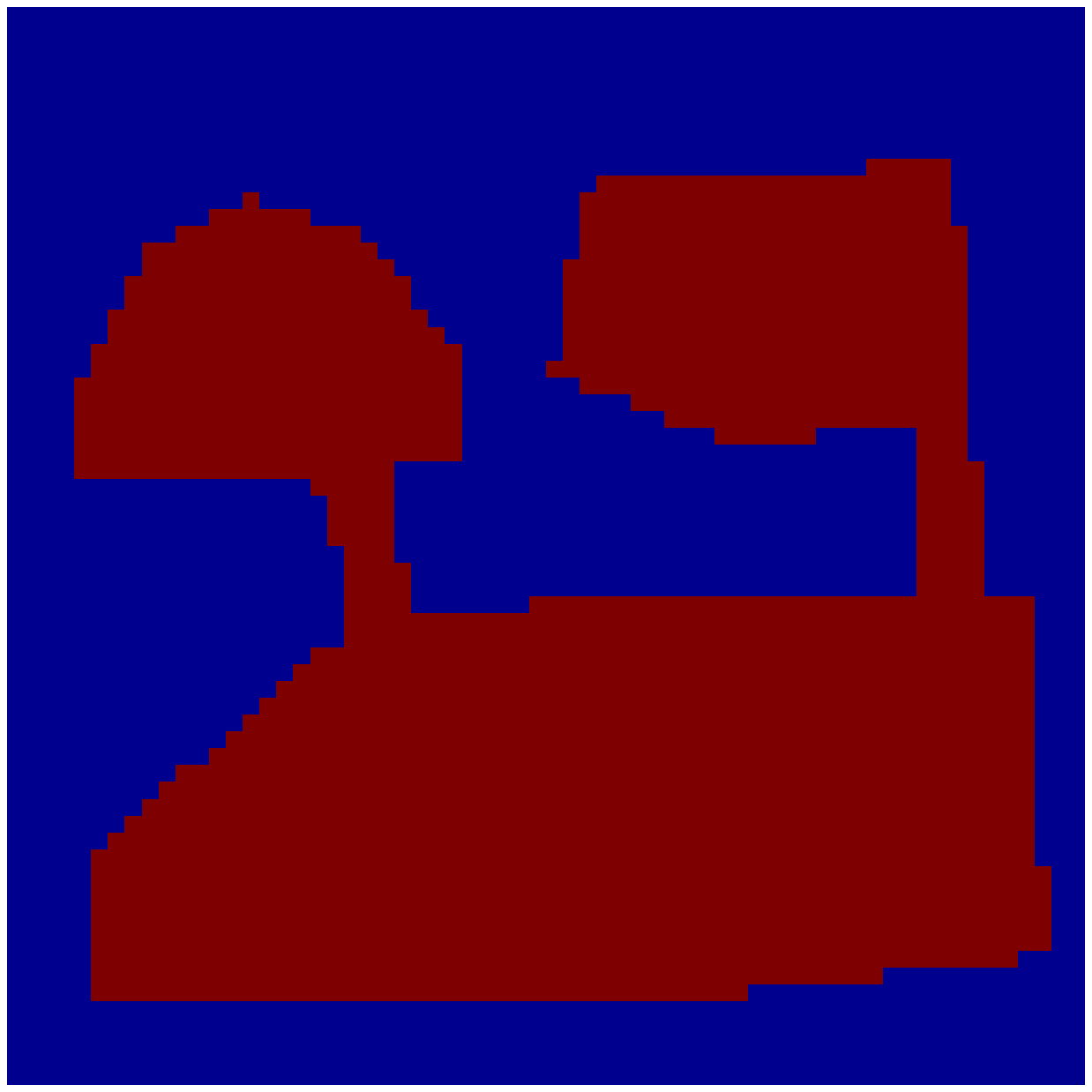} \\
\includegraphics[width=25mm,height=25mm]{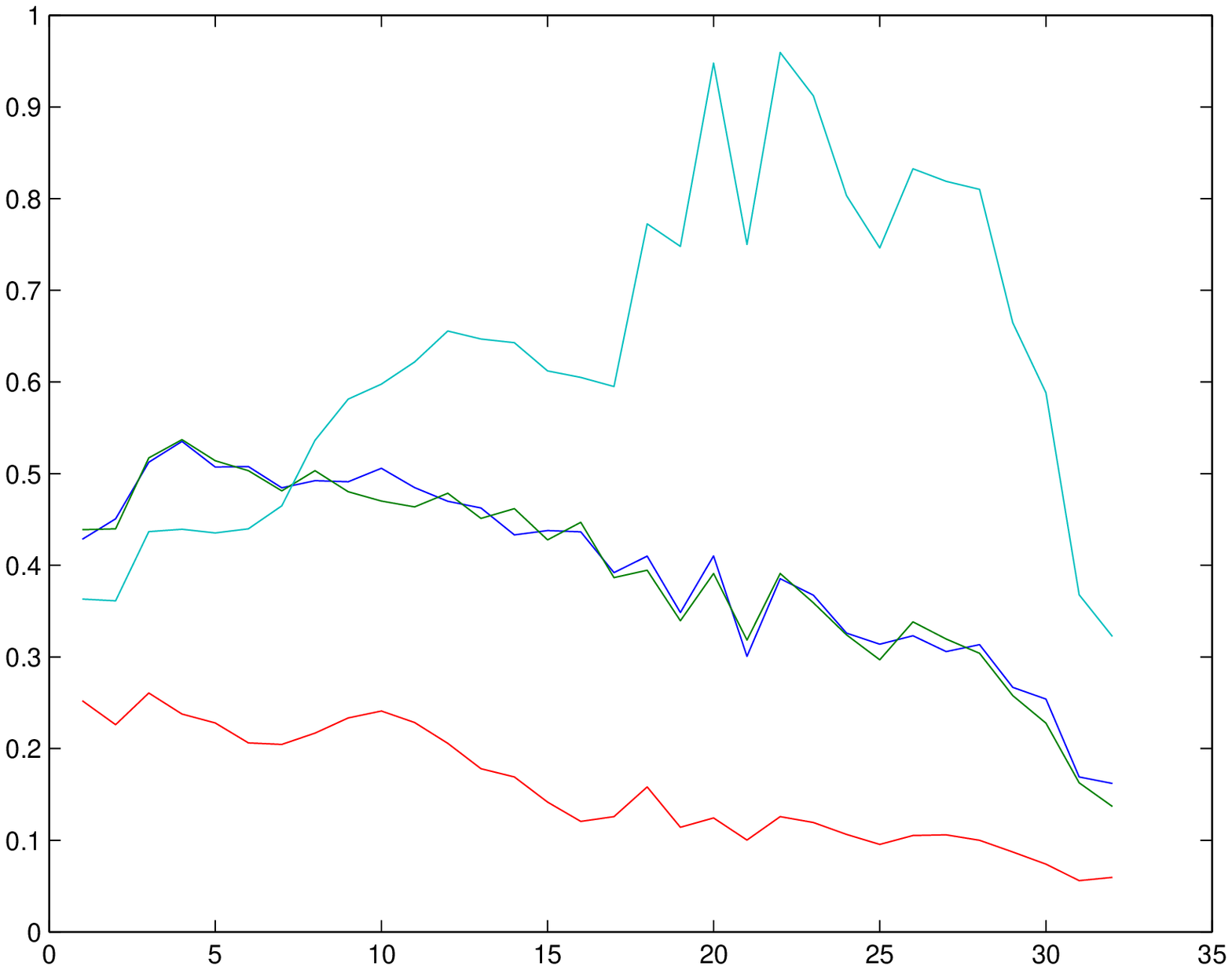}&
\includegraphics[width=25mm,height=25mm]{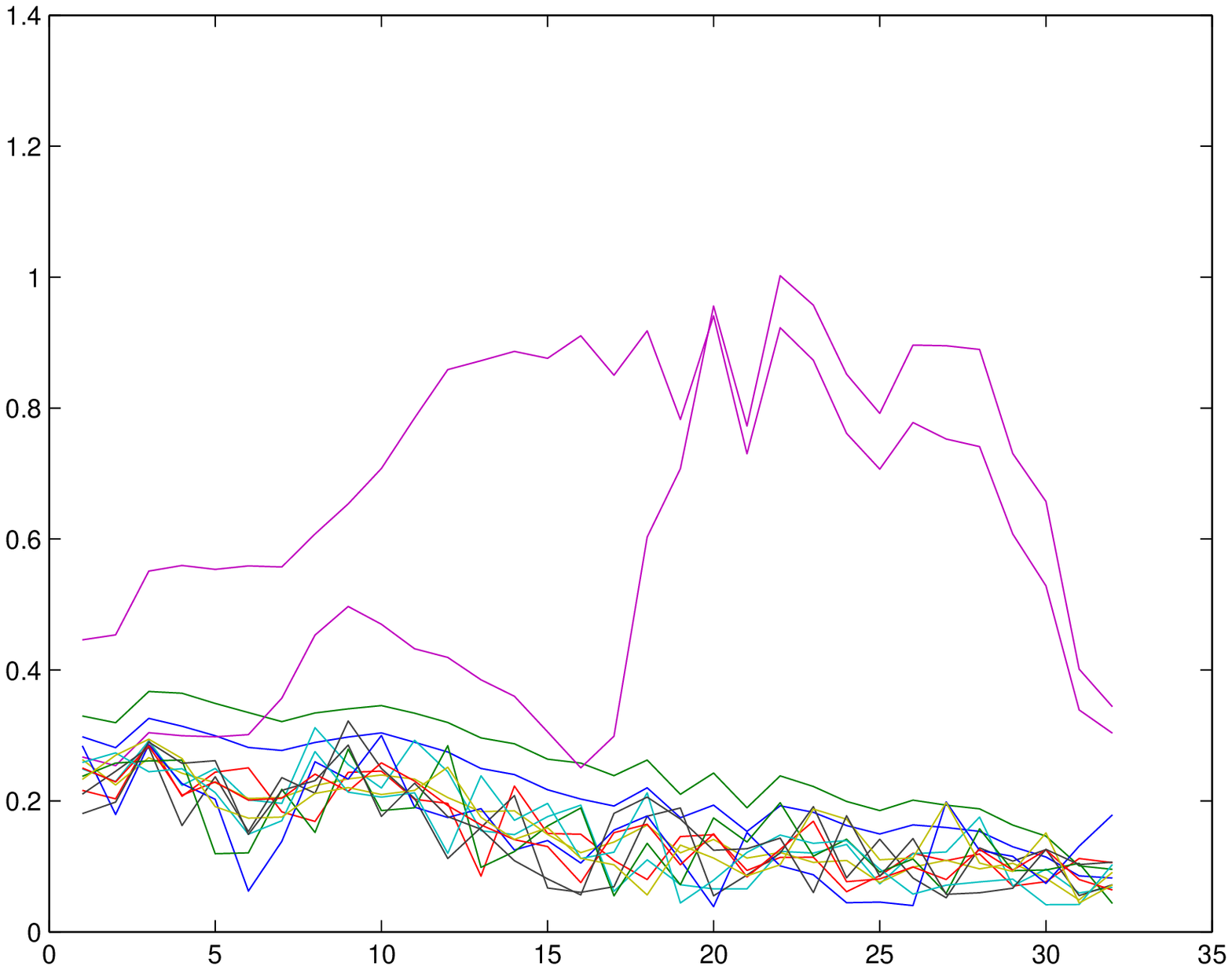}&
\includegraphics[width=25mm,height=25mm]{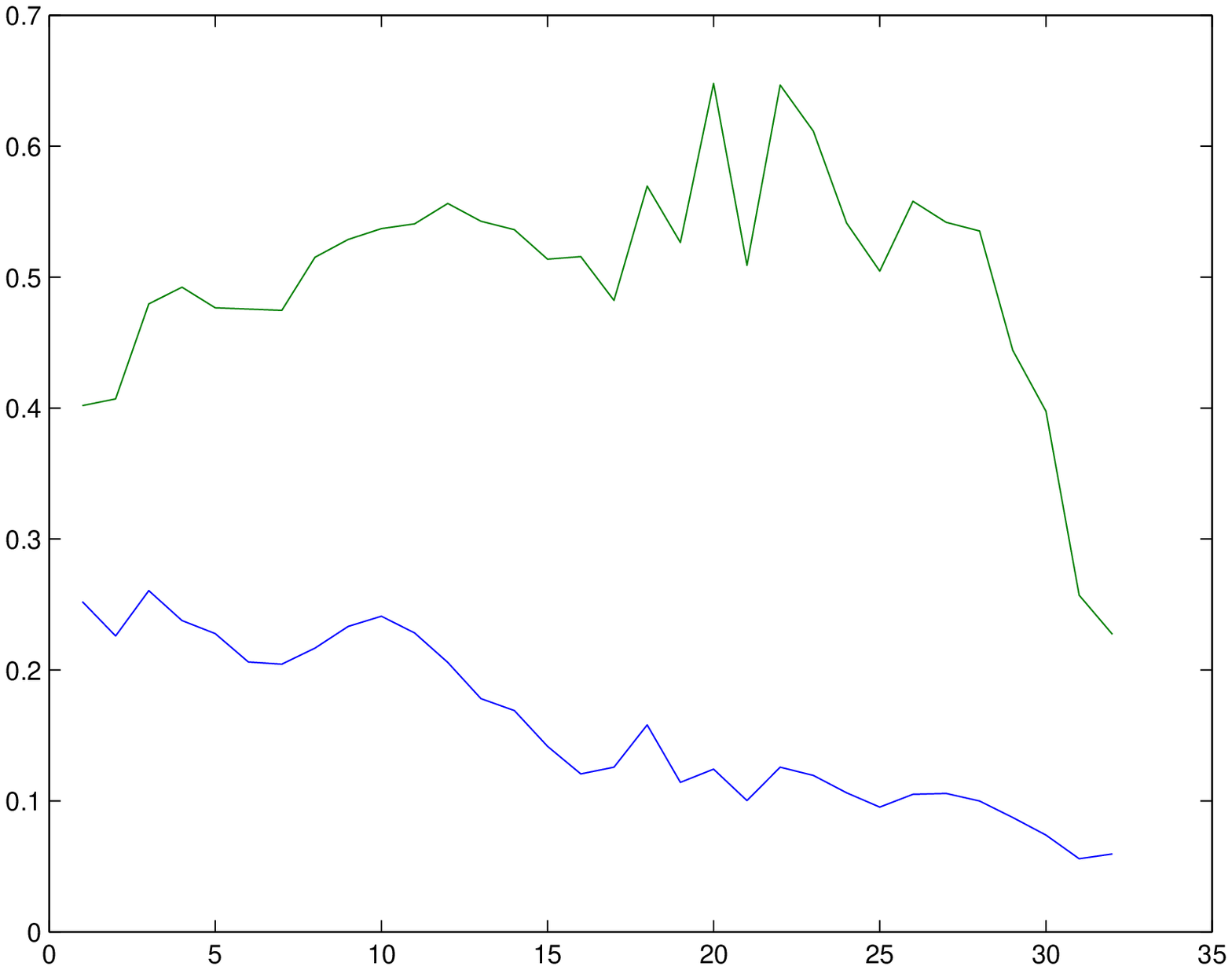} \\ 
a & b & c \\
\includegraphics[width=25mm,height=25mm]{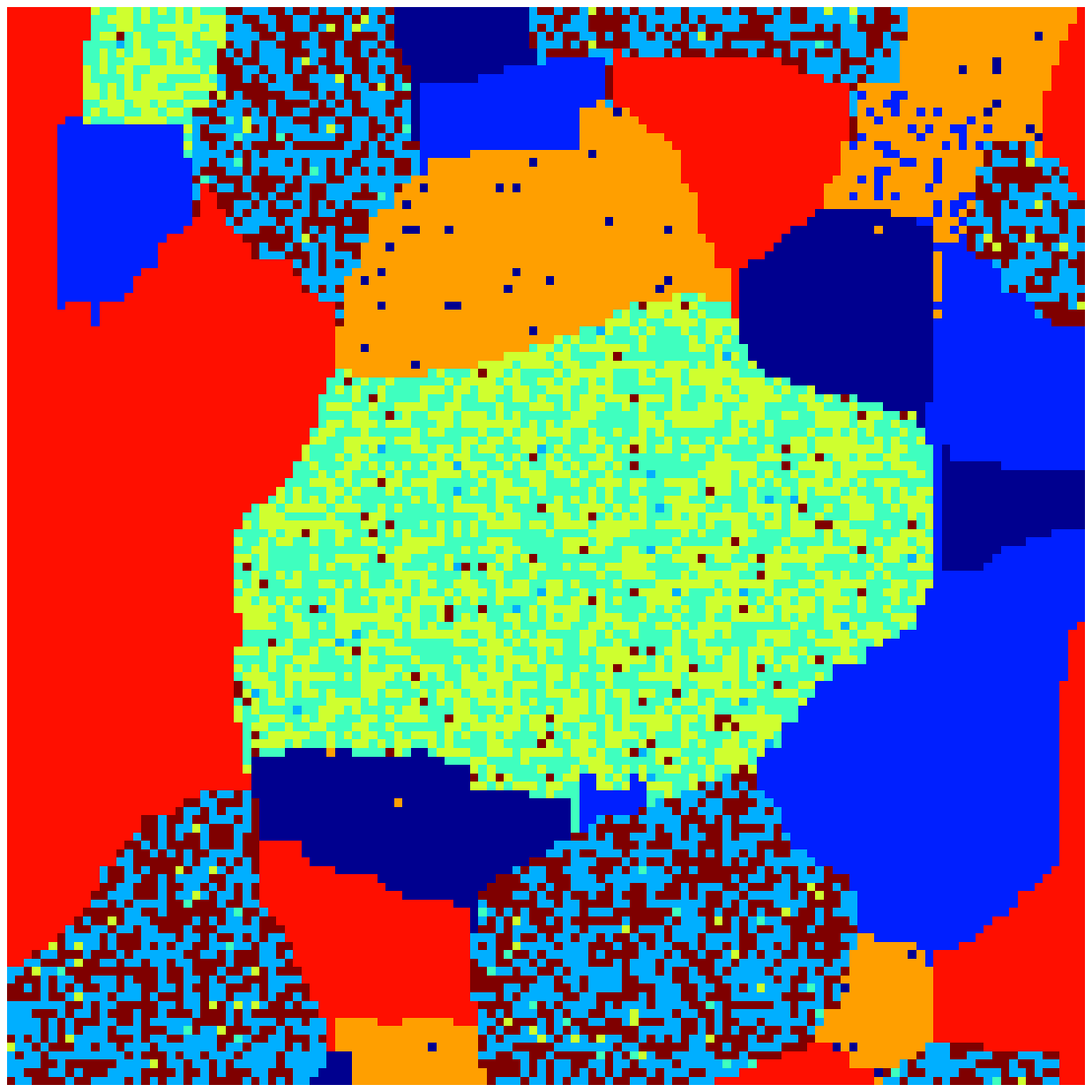}&
\includegraphics[width=25mm,height=25mm]{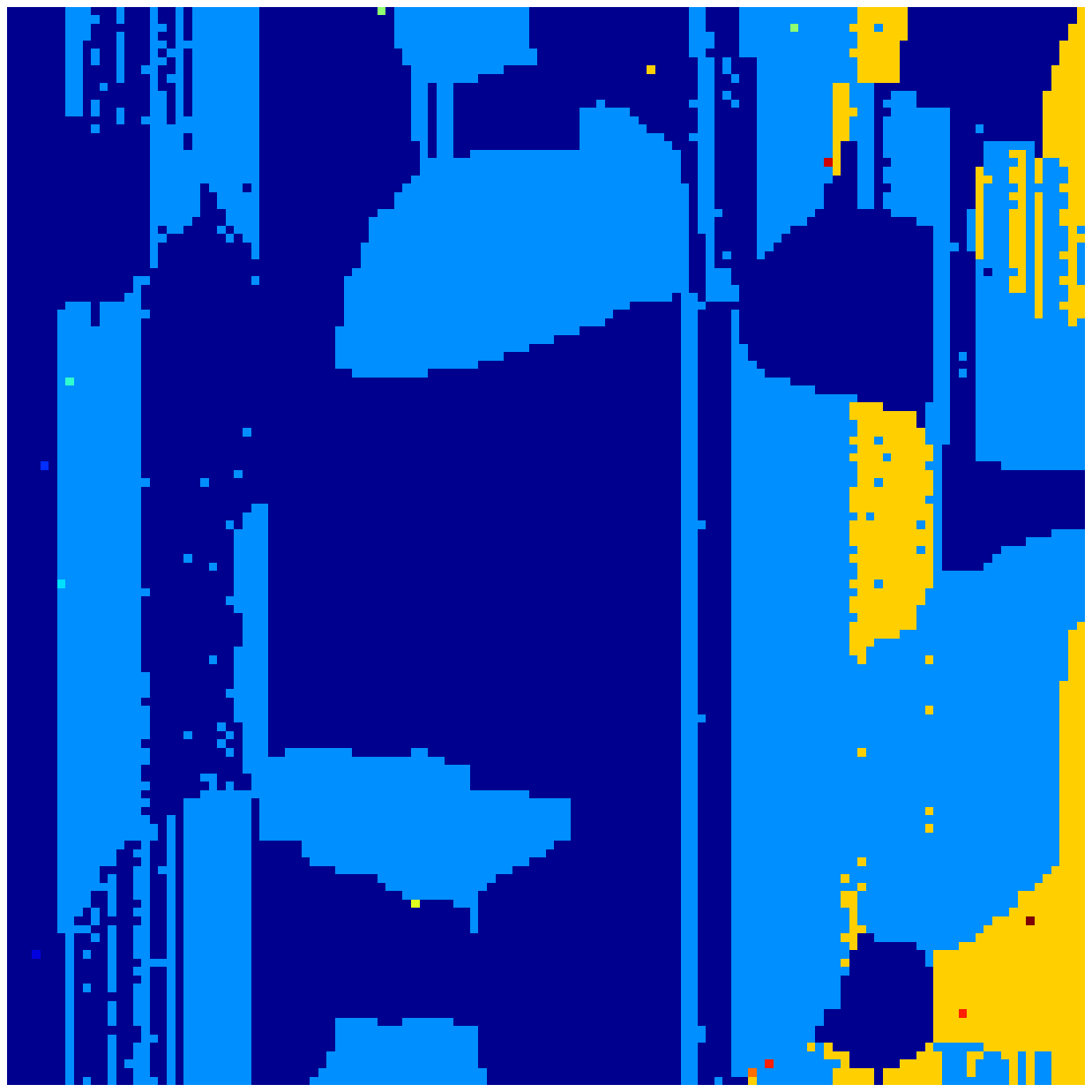}&
\includegraphics[width=25mm,height=25mm]{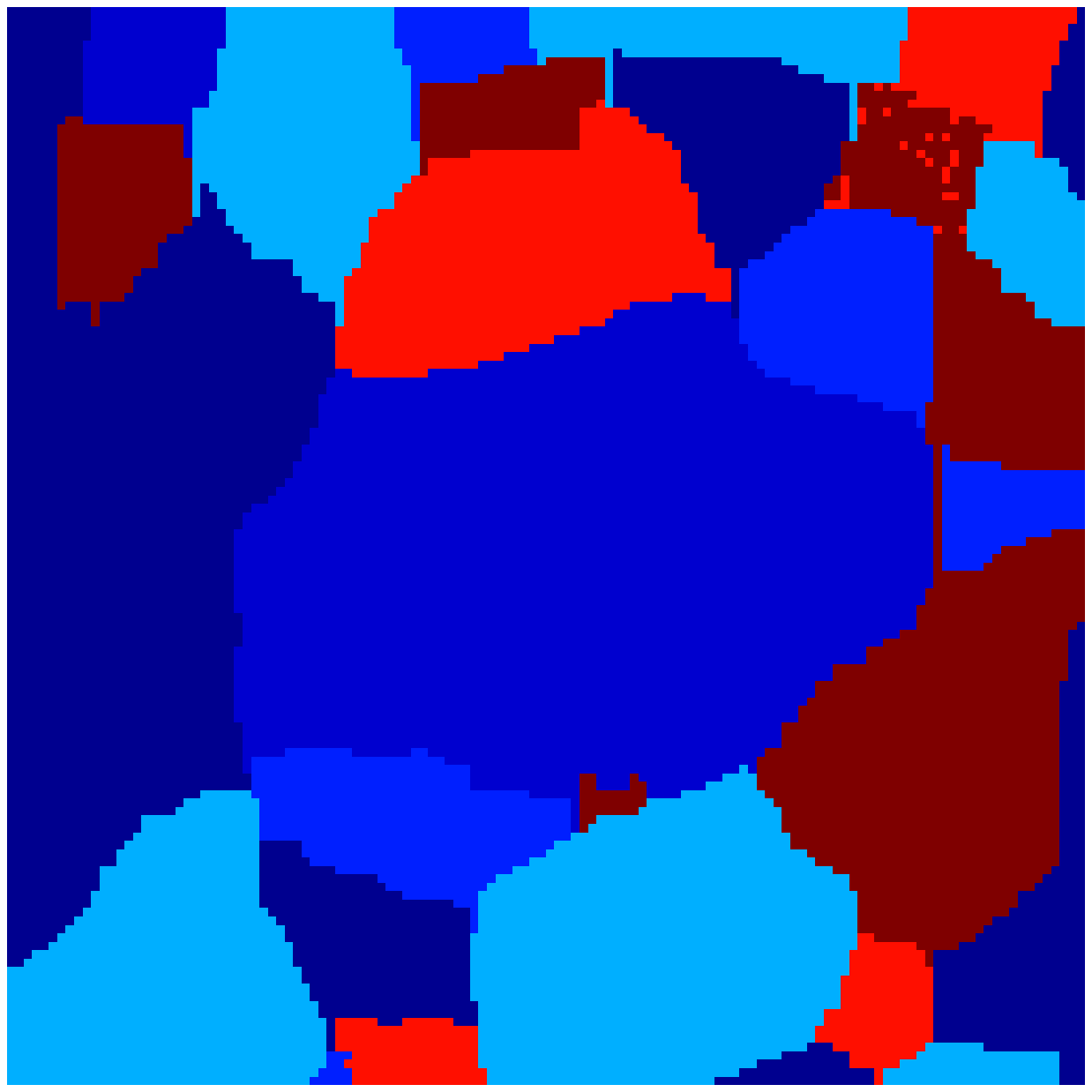} \\
\includegraphics[width=25mm,height=25mm]{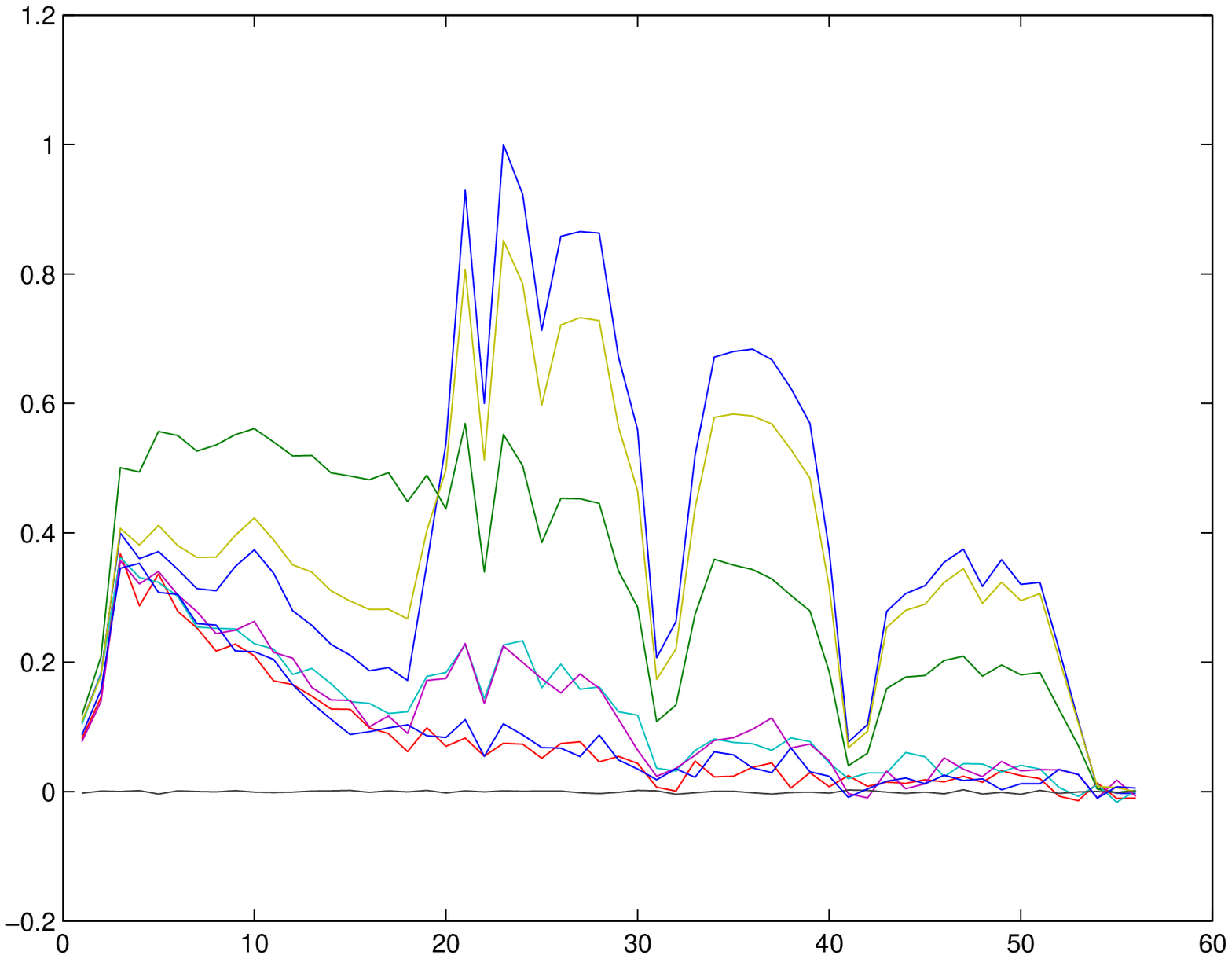}&
\includegraphics[width=25mm,height=25mm]{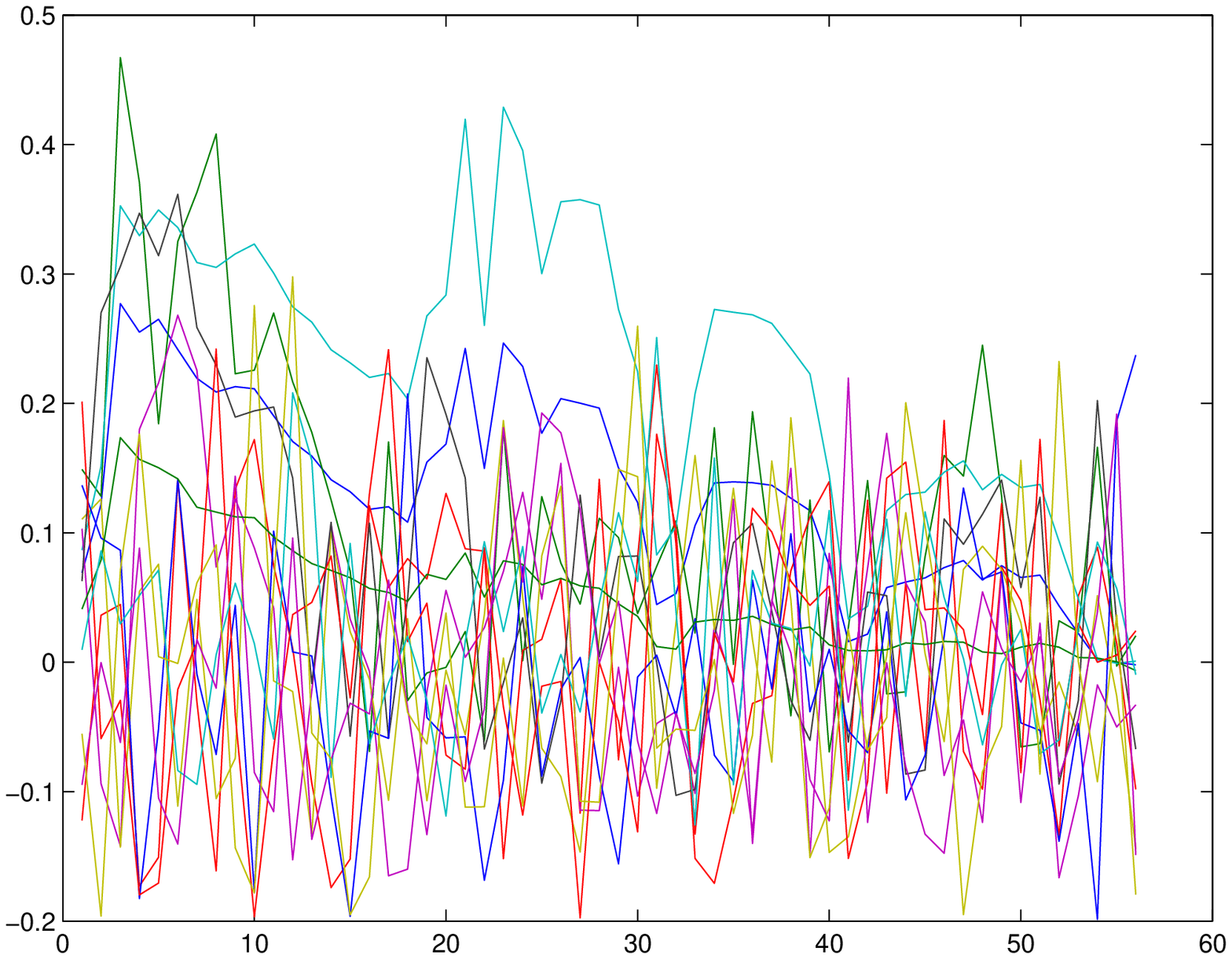}&
\includegraphics[width=25mm,height=25mm]{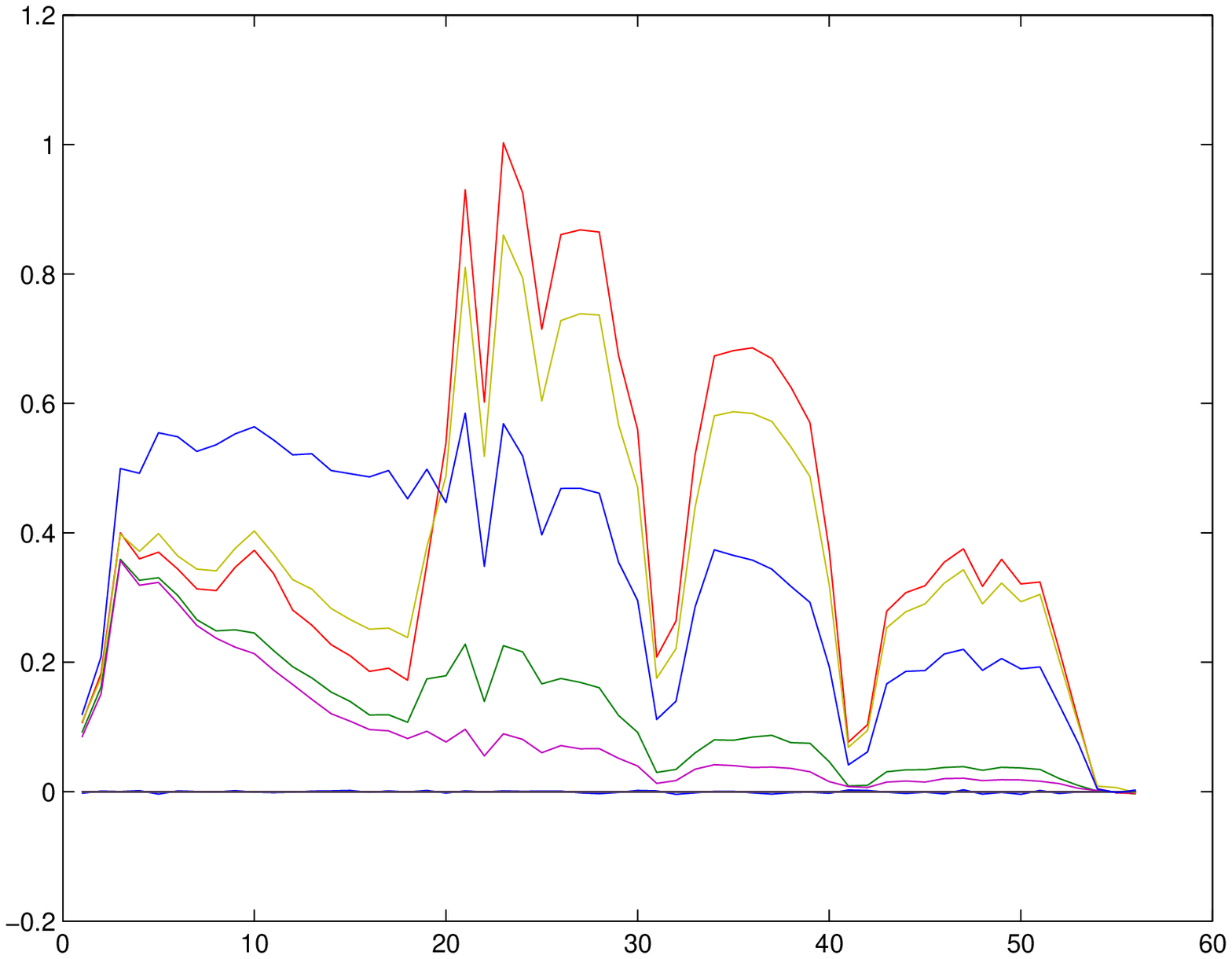} 
\etabu
\ecc
\caption{Dimensionality reduction by different methods: 
a) Spectral classification using $K$-means, 
b) Image classification using $K$-means, c) Proposed method. Upper row shows estimated $z(\rb)$ and lower row the estimated spectra. These results have to be compared to the original $z(\rb)$ and spectra in previous figure.}
\efig

\bfig[hbt]
\bcc
\btabu{@{}c@{}c@{}c@{}c@{}c@{}}
\includegraphics[width=25mm,height=25mm]{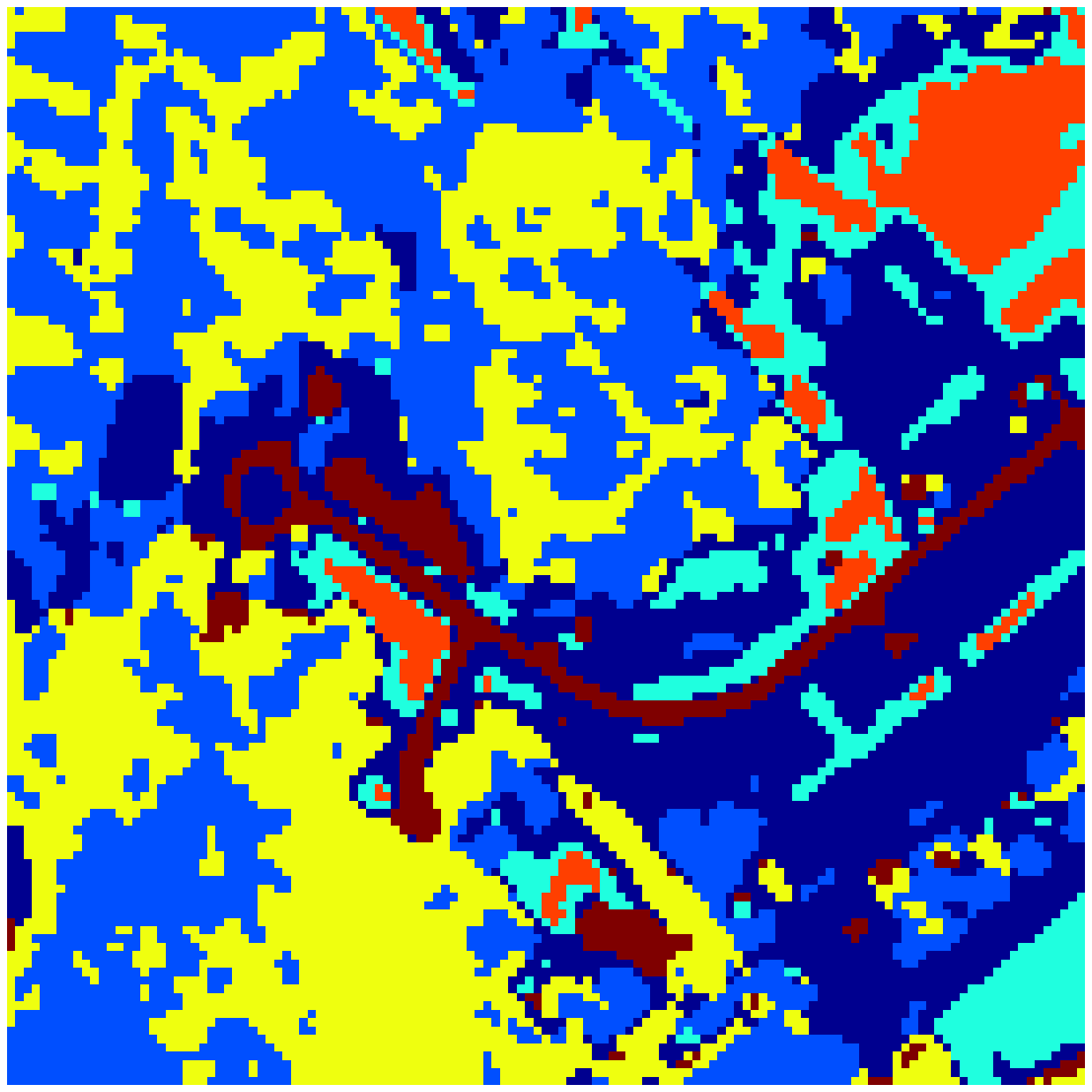}&
\includegraphics[width=25mm,height=25mm]{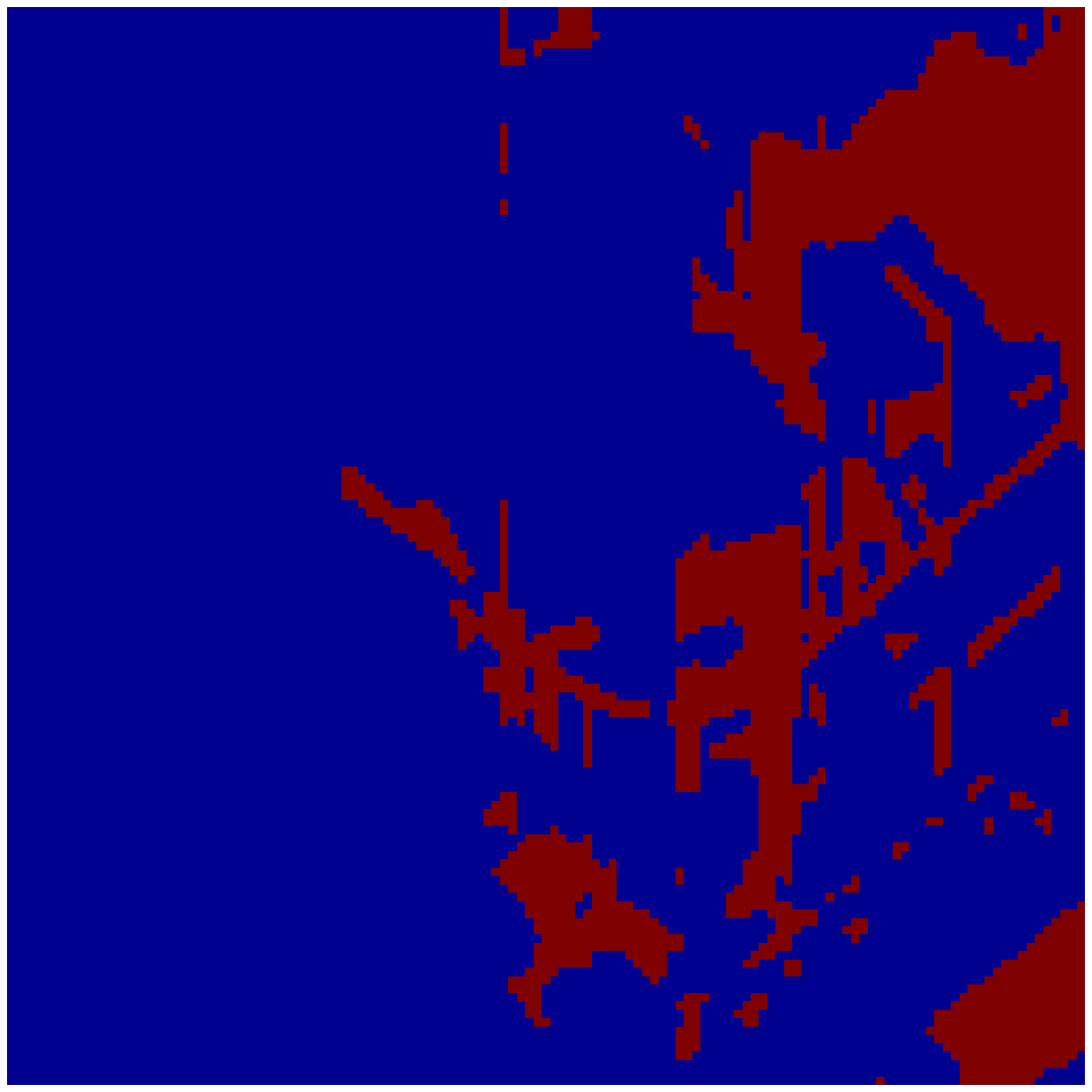}&
\includegraphics[width=25mm,height=25mm]{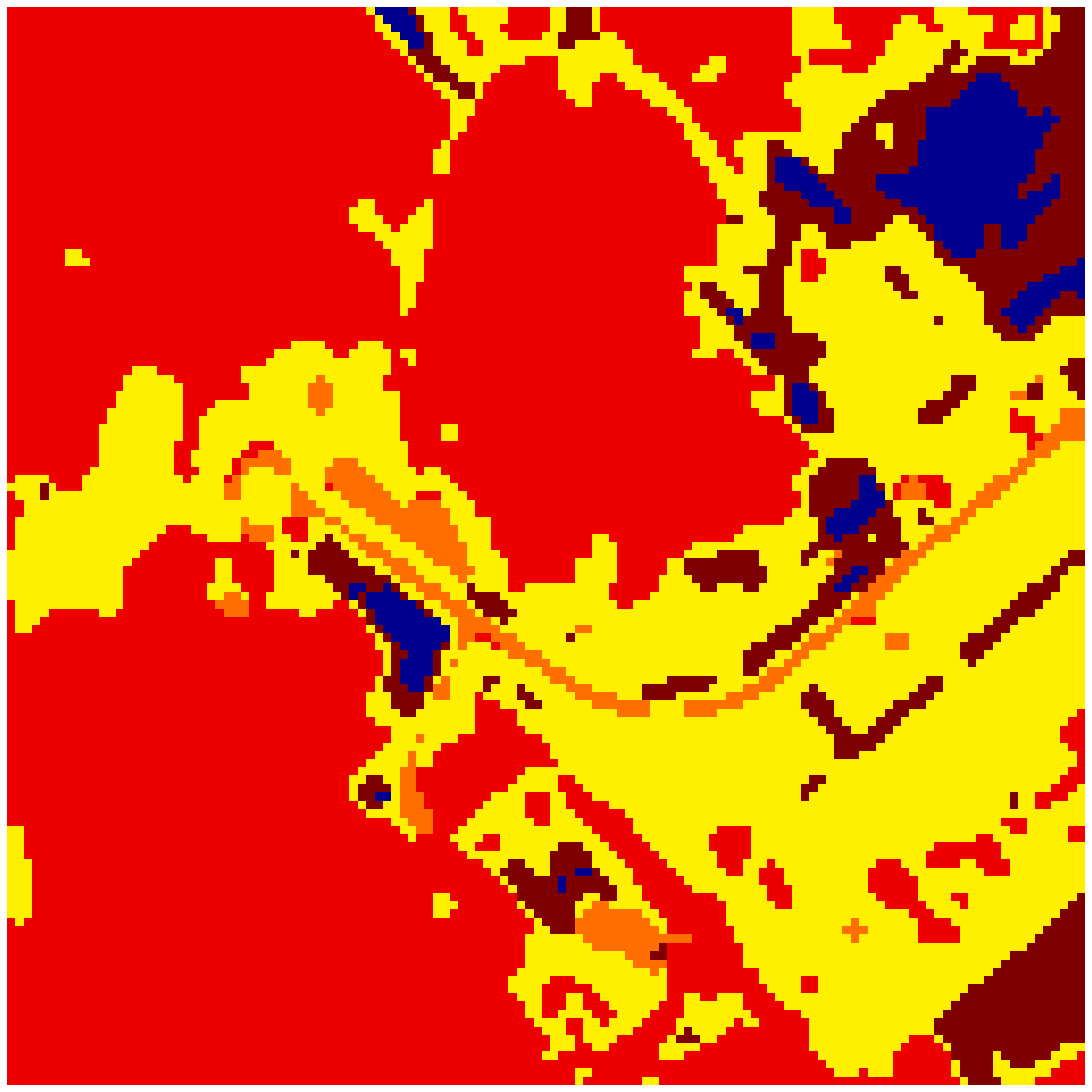} \\
\includegraphics[width=25mm,height=25mm]{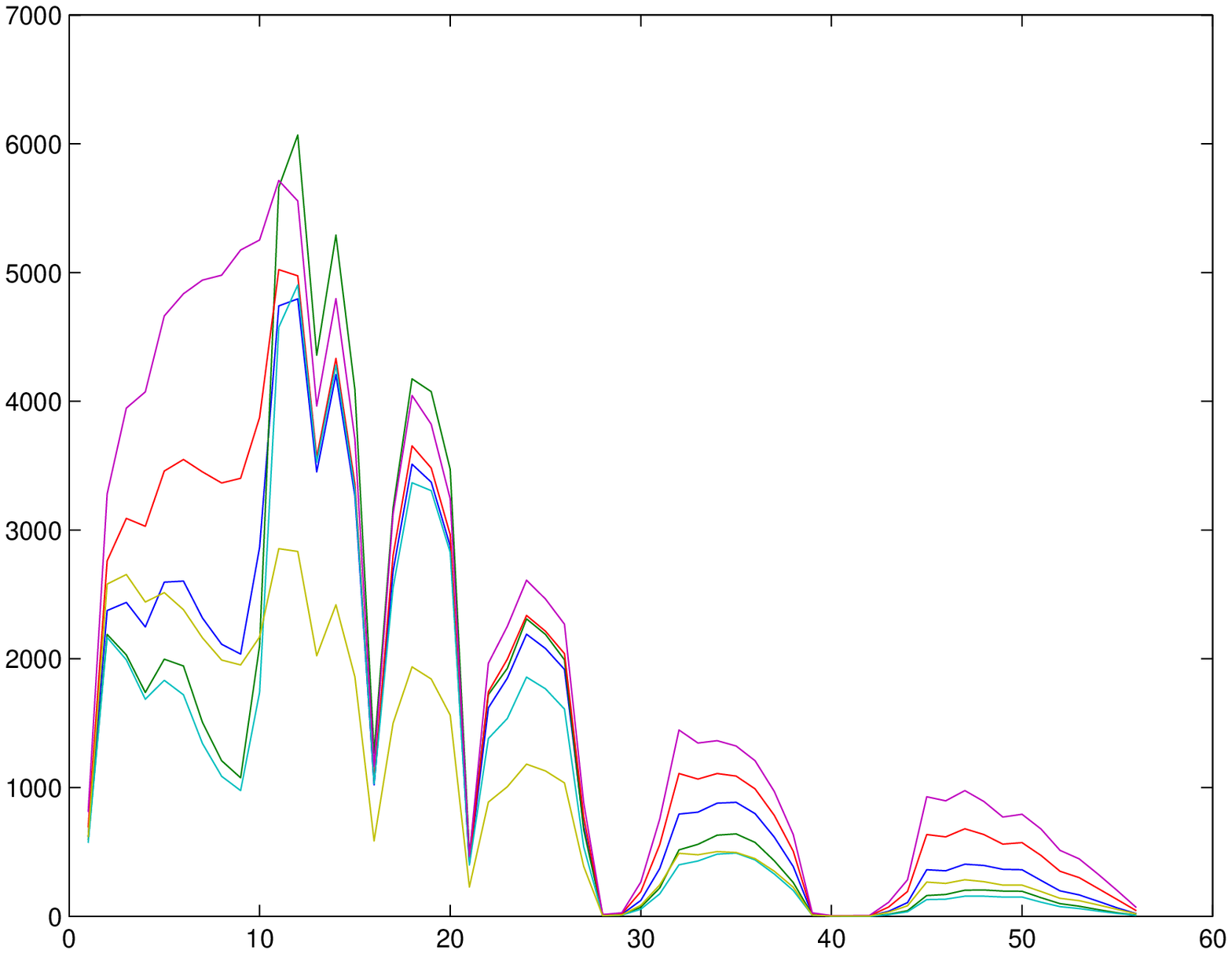}&
\includegraphics[width=25mm,height=25mm]{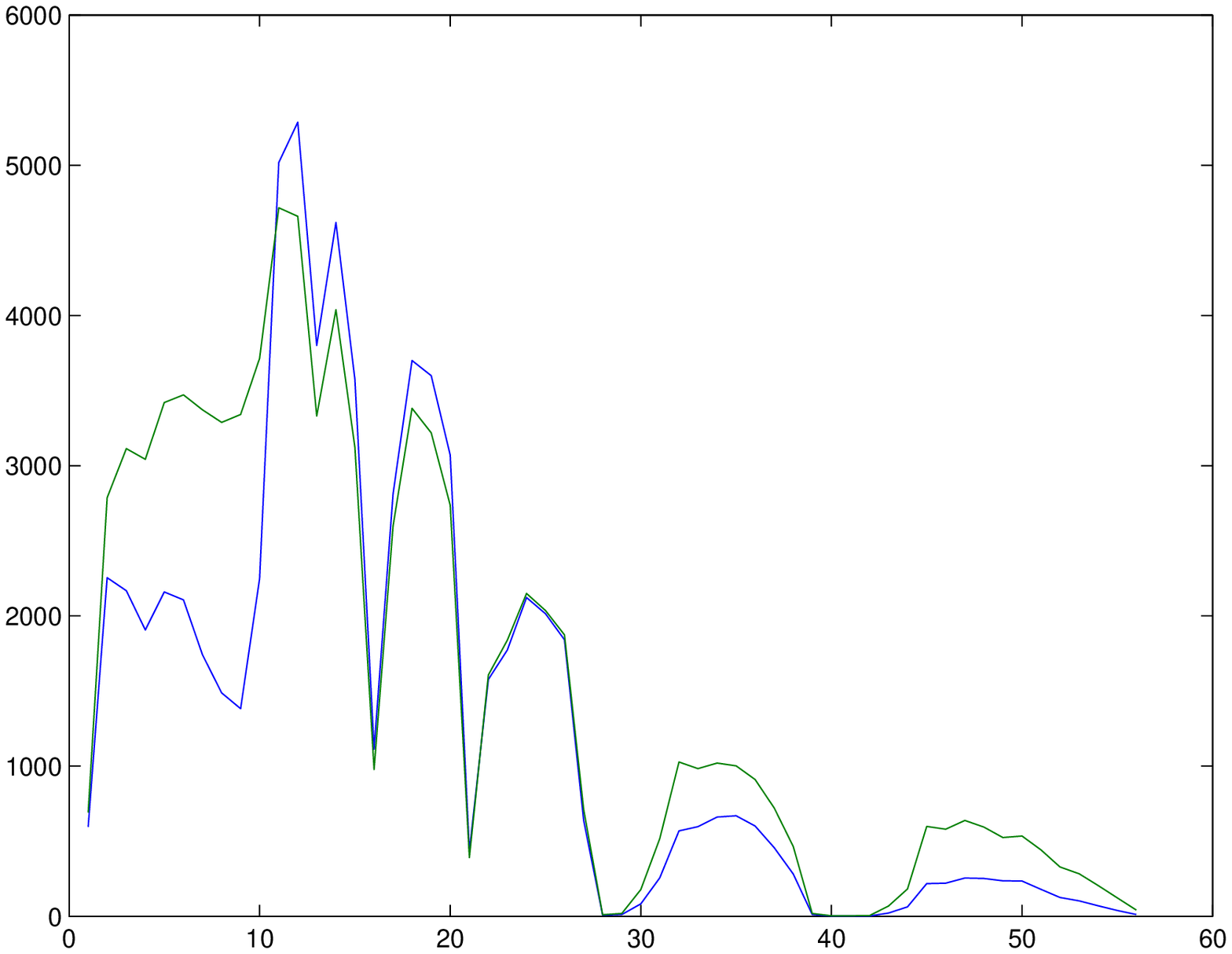}&
\includegraphics[width=25mm,height=25mm]{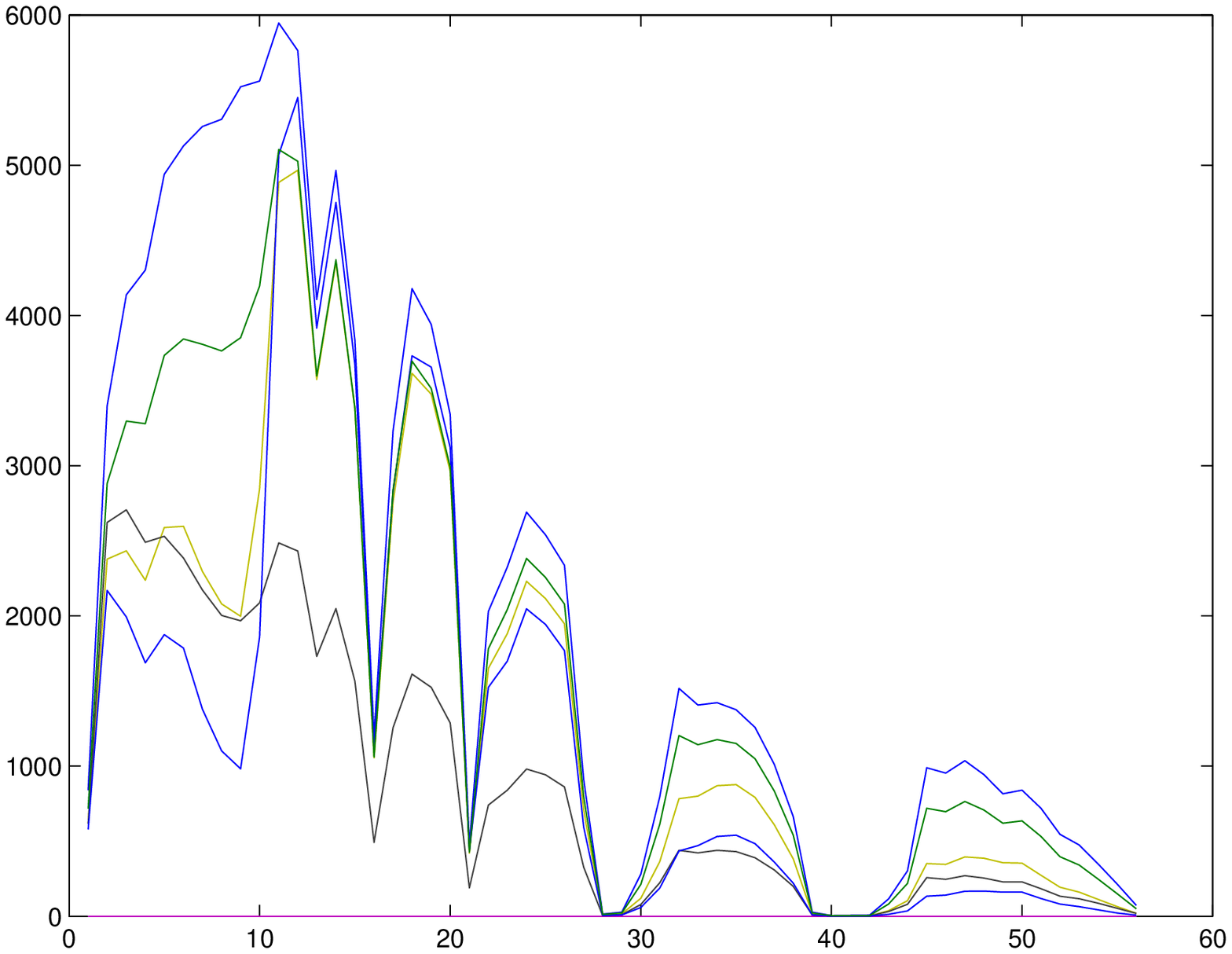} 
\etabu
\ecc
\caption{Real data: a) Spectral classification using $K$-means, 
b) Image classification using $K$-means, c) Proposed method. Upper row shows estimated $z(\rb)$ and lower row the estimated spectra.}
\efig

\vspace*{-6pt}
\section{Conclusion}
Classical methods of data reduction in hyperspectral imaging use classification methods either to classify the spectra or to classify the 
images in $K$ classes where $K$ is, in general, much less than the number of spectra or the number of observed images. However, these methods neglect 
either the spatial organization of the spectra or the spectral property of the pixels along the spectral bands. 
In this paper, we considered the dimensionality reduction problem in hyperspectral images as a source separation and presented a Bayesian estimation approach with an appropriate hierarchical prior model for the observations and sources which accounts for both spectral and spatial structure of the data, and thus, 
gives the possibility to jointly do dimensionality reduction, classification of spectra and segmentation of the images. 

\bibliographystyle{IEEEbib}
\def\bibdir{../Inputs/bib/}
\def\bibdirb{/home/djafari/Tex/Inputs/bib/amd/}
\def\sca#1{{\sc #1}}
{\small 
\bibliography{\bibdir bibenabr,\bibdir revuedef,\bibdir revueabr,\bibdir baseAJ,\bibdir baseKZ,\bibdir RapportThese,\bibdir gpipubli,\bibdirb amd_art,\bibdirb amd_ca,\bibdir icabook,adel}
}

\end{document}